\documentclass[fleqn,usenatbib]{mnras}

\usepackage{newtxtext,newtxmath}

\usepackage[T1]{fontenc}
\usepackage{ae,aecompl}

\usepackage{graphicx}	
\usepackage{amsmath}	
\usepackage{amssymb}	

\usepackage{bm}

\title[Efficient acceleration of cylindrical jets]{Efficient acceleration of cylindrical jets: effects of radiative cooling and tangled magnetic field}

\author[S. J. Tanaka \& K. Toma]{
Shuta J. Tanaka,$^{1}$\thanks{E-mail: sjtanaka@phys.aoyama.ac.jp (SJT)}
and Kenji Toma,$^{2,3}$
\\
$^{1}$Department of Physics and Mathematics, Aoyama Gakuin University, 5-10-1 Fuchinobe, Sagamihara, Kanagawa 252-5258, Japan \\
$^{2}$Frontier Research Institute for Interdisciplinary Sciences, Tohoku University, Sendai 980-8578, Japan\\
$^{3}$Astronomical Institute, Tohoku University, Sendai 980-8578, Japan
}

\date{Accepted XXX. Received YYY; in original form ZZZ}

\pubyear{2019}

\begin{document}
\label{firstpage}
\pagerange{\pageref{firstpage}--\pageref{lastpage}}
\maketitle

\begin{abstract}
	Diverging supersonic flows are accelerating, as in the case of a de Laval nozzle, and the same concept has been applied for acceleration of magnetohydrodynamic flows in the universe.
	Here, we study the dynamics of ``non-diverging'' cylindrical supersonic flows and show that they can be accelerated by effects of radiative cooling and the tangled magnetic field.
	In addition to radiative cooling of the jet materials (cooling effect), conversion of the ordered magnetic field into the turbulent one (conversion effect) and dissipation of the turbulent magnetic field (dissipation effect) are formulated according to our study on pulsar wind nebulae.
	Although each of the cooling and conversion effects is an ineffective acceleration process, the terminal velocity of magnetized cylindrical jets attains about half of the maximum possible value when the cooling, conversion and dissipation effects work simultaneously.
	The radiation efficiency is also about half of the total luminosity of the jet in the case of maximal acceleration.
	The concept for flow acceleration by the non-ideal MHD effects may be useful for studying relativistic jets in active galactic nuclei, in which the region near the jet axis is expected to be cylindrical and kink unstable.
\end{abstract}




\section{Introduction}\label{sec:intro}

Relativistic collimated outflows are observed from a variety of astrophysical systems, such as active galactic nuclei (AGNs), gamma-ray bursts, and microquasars.
Their engines are rotating compact objects and magnetic braking has been discussed as a mechanism of extracting their rotational energies.
In this scenario, the base of the jet should be Poynting dominated.
The broadband spectra of blazers, which are AGN jets pointing toward us, suggest that they are no longer Poynting dominated at $10^3 - 10^4 r_{\rm g}$ from their central supermassive black-holes, where $r_{\rm g}$ is the Schwarzschild radius \citep[e.g.,][]{Inoue&Takahara96, Ghisellini+10}.

Steady magnetohydrodynamic (MHD) flow has been studied for the acceleration mechanism of jets.
For a spherically symmetric radial flow, a Poynting dominated ideal MHD flow is not efficiently accelerated in contrast to a thermally dominated relativistic fireball.
Acceleration results from the magnetic and thermal pressure gradient in association with the flow expansion, while the magnetic tension acts against the magnetic pressure for the Poynting dominated flow.
\citet{Komissarov+07, Komissarov+09} found that axisymmetric ideal MHD flows confined by a rigid wall of a prescribed shape can be accelerated efficiently.
Because of self-collimation of the jet core, some poloidal magnetic field lines diverge faster than $B_{\rm p} \propto R^{-2}$ and then the flow is accelerated by converting the magnetic energy into the kinetic one, where $R$ is the cylindrical radius and $B_{\rm p}$ is the poloidal magnetic field strength \citep[cf.][]{Lyubarsky09, Beskin10, Toma&Takahara13}.
This relates with a well-known result that a cylindrical flow, i.e., a ``non-diverging'' flow, does not accelerate in the ideal MHD and even in the pure hydrodynamic (fireball) cases.

An alternative mechanism to accelerate a Poynting dominated flow is dissipation of the magnetic energy into the plasma heat.
In \citet{Lyubarsky&Kirk01} and \citet{Drenkhahn02}, the dissipation effect is added to the ideal MHD formulation in a phenomenological way, and then the dissipation time-scale is introduced as an additional parameter of the system.
For a specific choice of the dissipation time-scale, so called `striped wind model', the flow Lorentz factor increases as $\propto r^{1/3}$, which is not so efficient as the fireball model $\propto r$, where $r$ is the spherical radius.
The different exponent of the radial dependence of the flow Lorentz factor implies that the gradient of thermal pressure which originates from heating-up by magnetic dissipation is not a direct mechanism of flow acceleration.
The non-ideal MHD effect provides a different acceleration mechanism.

The above consideration suggests that the non-ideal MHD jet could be accelerated even for cylindrical geometry.
The recent VLBI radio observations of the M87 jet, which is one of the most extensively observed AGN jets, reveal that the overall jet shape is parabolic rather than cylindrical \citep[][]{Asada&Nakamura12, Hada+13, Nakamura+18}.
However, the cores of the jets are expected to be of cylindrical shape \citep[e.g.,][]{Beskin&Nokhrina09, Porth&Komissarov15}.
A hint of the jet core has also been observed from the same M87 jet, where an additional component between the bifurcated limb-brightened structure is reported \citep[][]{Asada+16, Hada17, Sobyanin17, Walker+18, Ogihara+19}.
The dynamics of cylindrical jets with the effects of the magnetic turbulence and radiative cooling is interesting for the application to AGN jets.

In this paper, we study flow acceleration based on the formulation developed by \citet{Tanaka+18} in cylindrical geometry, where the flow does not accelerate in the ideal MHD limit.
Our formulation is a simple extension of the magnetic dissipation model \citep[][]{Lyubarsky&Kirk01, Drenkhahn02} and includes cooling of the plasma, conversion of the ordered magnetic field into the turbulent one and dissipation of the turbulent magnetic field (section \ref{sec:Model}).
Radiative cooling is an essential property of the relativistic jets and recent numerical studies of relativistic jets also discuss the development of the turbulence and the magnetic dissipation inside the cylindrical core \citep[e.g.,][]{Porth&Komissarov15,Bromberg&Tchekhovskoy16, Bromberg+19}.
In section \ref{sec:AccelerationMechanisms}, the flow solutions for different acceleration mechanisms are described.
We find that a Poynting dominated cylindrical jet can be accelerated efficiently by the non-ideal MHD effects.
Overall discussion and the conclusions of the present paper are drawn in section \ref{sec:dis_cons}.

\section{Model}\label{sec:Model}

Our formulation of relativistic outflow dynamics with the turbulent magnetic field \citep{Tanaka+18} is applied to cylindrical jets.
Three non-ideal MHD terms are introduced in our formulation: (1) cooling of plasma, (2) conversion of the ordered toroidal magnetic field into the turbulent one, and (3) dissipation of the turbulent magnetic field into the heat of plasma.

For simplicity, we consider non-expanding pure cylindrical flows beyond the fast magnitosonic point.
This is because we focus on the non-MHD acceleration mechanisms rather than the ideal MHD acceleration associated with flow expansion (see also the discussion about the lateral force balance of cylindrical jets in section \ref{sec:dis_cons}).

\subsection{Basic equations}\label{sec:BasicEquations}

The assumptions to the flow properties are (i) steady, (ii) (cylindrical) one-dimension along the jet axis $z$, (iii) velocity field along the jet axis (no transverse component) and (iv) no lateral structure.
For the magnetic field, (v) transverse (toroidal) and turbulent components are considered.
The system has six variables: the proper enthalpy density $w$, the pressure $p$, the proper mass density $\rho$, the four velocity along the jet axis $u \equiv \gamma \beta$, the strength of the ordered transverse magnetic field $\bar{b}$ in proper frame and the strength of the turbulent magnetic field $\delta b$ which is isotropic in the proper frame ($\langle \delta \bm{b} \rangle = 0$ and $\langle \delta \bm{b}^2 \rangle = \delta b^2$).
See Appendix \ref{app:ModelOfNonIdealMHDEffects} for mathematical details of our formulation.

We adopt the equation of state of a Synge gas, i.e., a J\"uttner--Synge distribution plasma \citep{Juttner28, Synge57}.
Introducing the non-dimensional temperature $\Theta(z)$ which is normalized by the rest mass energy of the plasma particles, the enthalpy density and the pressure are described as
\begin{eqnarray}
	w(z)
	& = &
	\omega(\Theta) \rho c^2
	, \label{eq:EOS_enthalpy} \\
	p(z)
	& = &
	\Theta \rho c^2
	\label{eq:EOS_pressure}
\end{eqnarray}
with
\begin{eqnarray}
	\omega(\Theta)
	\equiv
	\frac{K_3(\Theta^{-1})}{K_2(\Theta^{-1})}, \label{eq:EOS_omega}
\end{eqnarray}
where $K_{\nu}(x)$ is the modified Bessel function of the second kind and $\omega(\Theta) \rightarrow 1$ for $\Theta \rightarrow 0$.
The internal energy density is written as
\begin{eqnarray}\label{eq:EOS_InternalEnergy}
	e_{\rm int}
	=
	w - p - \rho c^2
	=
	(\omega(\Theta) - \Theta - 1) \rho c^2.
\end{eqnarray}

Equation of continuity is immediately integrable and then the mass flux density,
\begin{eqnarray}\label{eq:MassFlux}
	\dot{m} \equiv \rho(z) u(z) c = {\rm const.},
\end{eqnarray}
is introduced.
The rest of the four equations are
\begin{eqnarray}
	\frac{d}{d z} \left[\gamma u \left(w + \bar{b}^2 + \frac{2}{3} \delta b^2 \right) \right]
	& = &
	- \gamma \frac{\Lambda_{\rm cool}}{c}
	, \label{eq:TotalEnergyConservation} \\
	\frac{d}{d z} (w - p) + w \frac{d}{d z} \ln u
	& = &
	\frac{\delta b^2 / 2}{u c \tau_{\rm diss}}
	- \frac{\Lambda_{\rm cool}}{u c}
	, \label{eq:InternalEnergyConservation} \\
	\frac{d}{d z} \frac{\bar{b}^2}{2} + \frac{\bar{b}^2}{2} \frac{d}{d z} \ln u^2
	& = &
	- \frac{\bar{b}^2 / 2}{u c \tau_{\rm conv}}
	, \label{eq:ToroidalMagneticEnergyConservation} \\
	\frac{d}{d z} \frac{\delta b^2}{2} + \frac{2}{3} \delta b^2 \frac{d}{d z} \ln u
	& = &
	- \frac{\delta b^2 / 2}{u c \tau_{\rm diss}}
	+ \frac{\bar{b}^2  / 2}{u c \tau_{\rm conv}}
	, \label{eq:TurbulentMagneticEnergyConservation}
\end{eqnarray}
which represent the conservation laws of the total energy flux, the plasma internal energy, and the energies of the ordered and turbulent magnetic field, respectively.
All the non-ideal MHD effects are on the right-hand side of equations (\ref{eq:TotalEnergyConservation}) $-$ (\ref{eq:TurbulentMagneticEnergyConservation}), where $\Lambda_{\rm cool}$, $\tau_{\rm conv}$ and $\tau_{\rm diss}$ are parameters of the system characterizing the cooling, conversion and dissipation effects, respectively.

\subsection{Differential equation for velocity}\label{sec:DifferentialEquationForVelocity}

For the sake of convenience, we introduce five quantities;
\begin{eqnarray}
	\epsilon(z)
	& \equiv &
	w + \bar{b}^2 + (2/3) \delta b^2
	, \label{eq:TotalEnergyDensity} \\
	l(z)
	& \equiv &
	\gamma u c \epsilon
	, \label{eq:TotalEnergyFlux} \\
	\sigma(z)
	& \equiv &
	\frac{\bar{b}^2 + (2/3)\delta b^2}{w}
	, \label{eq:Magnetization} \\
	\beta^2_{\rm c}(z)
	& \equiv &
	\frac{\hat{\Gamma} p + \bar{b}^2 + (2/9)\delta b^2}{\epsilon}
	, \label{eq:CharacteristicVelocity} \\
	\hat{\Gamma}(z)
	& \equiv &
	\frac{\omega'(\Theta)}{\omega'(\Theta) - 1}
	. \label{eq:EffectiveAdiabaticIndex}
\end{eqnarray}
The dash $'$ in equation (\ref{eq:EffectiveAdiabaticIndex}) denotes a derivative with respect to $\Theta$.
Equations (\ref{eq:TotalEnergyDensity}) $-$ (\ref{eq:EffectiveAdiabaticIndex}) represent the total energy (or enthalpy, precisely) density $\epsilon(z)$, the total energy flux $l(z)$, the magnetization $\sigma(z)$, the characteristic velocity $\beta_{\rm c}(z)$, and the effective adiabatic index $\hat{\Gamma}(z)$, respectively.
Combining equations (\ref{eq:TotalEnergyConservation}) $-$ (\ref{eq:TurbulentMagneticEnergyConservation}) with the use of the above quantities, we obtain the differential equation for the four velocity
\begin{eqnarray}\label{eq:DifferentialEquationForVelocity}
	\epsilon (\beta^2 - \beta^2_{\rm c}) \frac{d u}{d z}
	& = &
	(\hat{\Gamma} - 1) \frac{\Lambda_{\rm cool}}{c} \nonumber \\
	& + &
	\frac{\bar{b}^2}{3 c \tau_{\rm conv}}
	+
	\left(\frac{4}{3} - \hat{\Gamma} \right) \frac{\delta b^2}{2 c \tau_{\rm diss}}.
\end{eqnarray}
Equation (\ref{eq:DifferentialEquationForVelocity}) corresponds to the equation (19) of \citet{Tanaka+18} although there are two important differences.

First, in the ideal MHD limit ($\Lambda_{\rm cool} = 0$ and $\tau_{\rm conv},\tau_{\rm diss} \rightarrow \infty$), we obtain the well-known conclusion that the cylindrical jet does not accelerate ($d u / d z = 0$) because equation (\ref{eq:DifferentialEquationForVelocity}) does not have the (geometrical) `expansion acceleration' term which is the first term of the right-hand side of the equation (19) of \citet{Tanaka+18}, and is proportional to $p / r$.
Second, the effective adiabatic index $\hat{\Gamma}$ ranges from 4/3 (ultra-relativistic temperature $\Theta \gg 1$) to 5/3 (non-relativistic temperature $\Theta \ll 1$) for a Synge gas, while we set a constant adiabatic index of 4/3 (i.e., a relativistically-hot gas) in our past study.
Only the last term of the right-hand side of equation (\ref{eq:DifferentialEquationForVelocity}) is negative for a Synge gas, i.e., they work as flow deceleration for $\beta > \beta_{\rm c}$.

\subsection{Numerical implementation}\label{sec:NumericalImplementation}

We solve the five differential equations (\ref{eq:MassFlux}) $-$ (\ref{eq:TurbulentMagneticEnergyConservation}) with the help of the equation of state (equations (\ref{eq:EOS_enthalpy}) $-$ (\ref{eq:EOS_omega})).
The corresponding five variables are $\rho(z) c^2,~u(z),~\Theta(z),~\bar{b}^2(z),$ and $\delta b^2(z)$.
Five boundary conditions are set at an inlet boundary $z = z_0$ and we use a normalized position $x \equiv z / z_0$ below.
It should be noted that we set $z_0 > 0$ and also $u(z) > 0$ only for simplicity while setting $z_0 \le 0$ does not change any discussion below.
The cylindrical axis $z$ (or $x$) can be zero and also negative unlike the spherical radius $r$.
The inlet boundary $z_0$ ($x = 1$) is not related with the Schwarzschild `radius' at all (see also section \ref{sec:dis_cons}).

In this study, three of the five inlet boundary conditions are fixed: (I) $l_0 / c = \gamma_0 u_0 \omega(\Theta_0) \rho_0 c^2 (1 + \sigma_0) = 1$, (II) $\gamma_{\rm max} \equiv l_0 / (\dot{m} c^2) = 10^4$, and (III) $\delta b^2_0 = 0$.
From the boundary condition (I), the total energy flux is the same for all the cases and the energy densities, $\rho(z) c^2, \bar{b}^2(z),$ $\delta b^2(z)$ and so on, are normalized by $l_0 / c = 1$, i.e., they are non-dimensional below.
The boundary condition (II) fixes the maximum attainable Lorentz factor to be the same for all the flows.
The boundary condition (III) is the assumption that the magnetic field is totally ordered at the inlet and then another boundary condition $\sigma_0 = \bar{b}^2_0 / (\omega(\Theta_0) \rho_0 c^2)$ is required to give the magnetization at the inlet.
In section \ref{sec:AccelerationMechanisms}, we study both high- and low-$\sigma_0$ cases by setting (IV) $\sigma_0 = 10^{2}$ and $10^{-2}$, respectively.

For the last inlet boundary condition, the four velocity is set to be the characteristic velocity (equation (\ref{eq:CharacteristicVelocity})), (V) $u_0 = u_{\rm c,0}$, where
\begin{eqnarray}\label{eq:InletVelocity}
	u^2_{\rm c,0}
	& = &
	\frac{(\hat{\Gamma}-1+\sigma_0) \gamma_{\rm max} - (\hat{\Gamma}-1)(1+\sigma_0)\gamma_0}
	{(2-\hat{\Gamma}) \gamma_{\rm max} + (\hat{\Gamma}-1)(1+\sigma_0)\gamma_0}.
\end{eqnarray}
The inlet boundary is not arbitrarily point anymore and is the `trans-characteristic-velocity' point ($u_0 = u_{\rm c,0}$).
In practice, the inlet four velocity $u_0$ is infinitesimally larger than $u_{\rm c,0}$ in order to obtain the acceleration solutions (see equation (\ref{eq:DifferentialEquationForVelocity})).

In conclusion, we study only two ($\sigma_0 = 10^2$ and $10^{-2}$) cases at the inlet boundary $z = z_0$ ($x = 1$).
Some results with the other $\sigma_0$ are briefly shown in appendix \ref{app:VariousInletBoundary}.
From $l_0 / c = 1$, all the results shown in the next section \ref{sec:AccelerationMechanisms} are dimensionless.
In order to relate our results with the power of the jet in real systems, in addition to $l_0$, we need to specify the cylinder (jet) cross section.
There still remains the arbitrariness of the origin of coordinates or of the length scale along the cylinder $z_0$ which is the distance between the origin and the trans-characteristic-velocity point.
It should be noted that, from the inlet boundary conditions (I) and (II), the boundary conditions must satisfy a physical constraint $\gamma_{\rm max} \ge \gamma_0 (1 + \sigma_0)$ for $\Theta_0 \ge 0$ ($\omega(\Theta_0) \ge 0$).
The constraint gives an upper limit of $\sigma_0$ because $\gamma_0 = \gamma_{\rm c} \sim \sqrt{\sigma_0}$ for $\sigma_0 \gg 1$, i.e., $\sigma_0 \lesssim \gamma_{\rm max}^{2/3}$.

We have three parameters of the system.
They characterize the non-ideal MHD effects.
For the conversion and dissipation effects, the normalized length-scales $x_{\rm conv} \equiv c \tau_{\rm conv} / z_0$ and $x_{\rm diss} \equiv c \tau_{\rm diss} / z_0$ are introduced as the parameters.
For the cooling effect, we set the form of the cooling term as
\begin{eqnarray}\label{eq:CoolingEffect}
	\frac{\Lambda_{\rm cool}}{c} z_0
	& = &
	\frac{e_{\rm int}}{x_{\rm cool}}
\end{eqnarray}
with the use of the internal energy density of plasma $e_{\rm int}$ \citep[cf.][]{Drenkhahn&Spruit02}.
Acceleration of cylindrical jets is studied for $x_i = 10, 10^3, 10^5$ and $\infty$ ($i =$ cool, conv and diss) by setting the outer boundary of $x_{\rm out} = 10^{10}$.

\section{Acceleration Mechanisms}\label{sec:AccelerationMechanisms}

We focus on flow acceleration ($\beta > \beta_{\rm c}$).
The unit $c = 1$ is adopted below.
The first two terms on the right-hand side of equation (\ref{eq:DifferentialEquationForVelocity}), i.e., the cooling and conversion effects, are important in this study, while the dissipation effect is subdominant, i.e., the flow has little chance to decelerate (sections \ref{sec:CoolingAcceleration}, \ref{sec:ConversionAcceleration}).
The dissipation effect heats up the plasmas and then it contributes indirectly to flow acceleration through the cooling effect rather than deceleration (section \ref{sec:EfficientAcceleration}).

Flow acceleration by the cooling and conversion effects is studied separately in sections \ref{sec:CoolingAcceleration} and \ref{sec:ConversionAcceleration}, respectively.
In section \ref{sec:EfficientAcceleration}, we study the case when all the three non-ideal MHD effects work.
The flows experience both conversion and cooling acceleration and are efficiently accelerated close to $\gamma_{\rm max}$ in some cases.
The adopted parameters and the derived terminal values for each result in this section are summarized in Table \ref{tbl:parameters}.

\subsection{Cooling acceleration}\label{sec:CoolingAcceleration}

%
\begin{table}
	\centering
	\caption{
		Summary of the adopted inlet boundary condition ($\sigma_0$), the characteristic length-scales ($x_{\rm cool}, x_{\rm conv}, x_{\rm diss}$) and the derived terminal values ($u_{\infty}, \sigma_{\infty}, l_{\infty}$) for Figs. \ref{fig:LowSigma} -- \ref{fig:Both}.
	}
	\label{tbl:parameters}
	\begin{tabular}{ccccccc} 
		\hline
		\multicolumn{4}{c}{Adopted} &
		\multicolumn{3}{c}{Derived} \\
		\hline
		$\sigma_0$ & $x_{\rm cool}$ & $x_{\rm conv}$ & $x_{\rm diss}$ & $u_{\infty}$ & $\sigma_{\infty}$ & $l_{\infty}$ \\
		\hline
		\multicolumn{7}{c}{Fig. \ref{fig:LowSigma}} \\
		$10^{-2}$ & $10    $ & $\infty$ & $\infty$ & 32   & 1.8 & 0.0089 \\
		$10^{-2}$ & $10^{3}$ & $\infty$ & $\infty$ & 32   & 1.8 & 0.0089 \\
		$10^{-2}$ & $10^{5}$ & $\infty$ & $\infty$ & 32   & 1.8 & 0.0089 \\
		$10^{-2}$ & $10^{3}$ & $10^{3}$ & $\infty$ & 13   & 11  & 0.017  \\
		$10^{-2}$ & $10^{3}$ & $10^{3}$ & $10^{3}$ & 32   & 0   & 0.003  \\
		\multicolumn{7}{c}{Fig. \ref{fig:Conversion}} \\
		$10^{ 2}$ & $\infty$ & $10    $ & $\infty$ & 150  & 17  & 1      \\
		$10^{ 2}$ & $\infty$ & $10^{3}$ & $\infty$ & 150  & 17  & 1      \\
		$10^{ 2}$ & $\infty$ & $10^{5}$ & $\infty$ & 150  & 17  & 1      \\
		$10^{ 2}$ & $\infty$ & $10    $ & $10    $ & 150  & 0   & 1      \\
		$10^{ 2}$ & $\infty$ & $10^{3}$ & $10    $ & 150  & 0   & 1      \\
		$10^{ 2}$ & $\infty$ & $10^{5}$ & $10    $ & 150  & 0   & 1      \\
		\multicolumn{7}{c}{Fig. \ref{fig:Both}} \\
		$10^{ 2}$ & $10    $ & $10    $ & $10    $ & 790  & 0   & 0.079  \\
		$10^{ 2}$ & $10    $ & $10^{3}$ & $10    $ & 4700 & 0   & 0.47   \\
		$10^{ 2}$ & $10    $ & $10^{5}$ & $10    $ & 5000 & 0   & 0.5    \\
	\end{tabular}
\end{table}
\begin{figure}
	\includegraphics[width=\columnwidth]{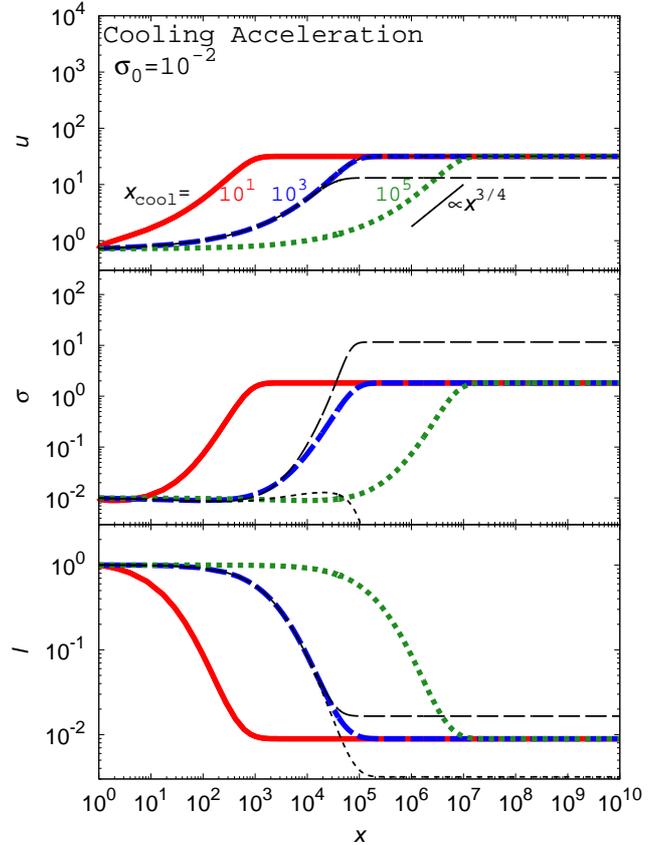}
    \caption{
	    The profiles of the four velocity $u(x)$, the magnetization $\sigma(x)$ and the total energy flux $l(x)$ when cooling acceleration is dominant.
    The flows are a relativistically-hot low-$\sigma_0$ flow at the inlet boundary $x = 1$ ($\sigma_0 = 10^{-2}$).
    The conversion and dissipation effects are ignored for thick lines ($x_{\rm conv} = x_{\rm diss} = \infty$) and $x_{\rm cool}$ is 10 (thick red solid lines), 10$^3$ (thick blue dashed lines), and 10$^5$ (thick green dotted lines), respectively.
    For thin black lines, $x_{\rm cool} = x_{\rm conv} = 10^3$ is common while $x_{\rm diss}$ is 10 (thin black dotted lines) and $\infty$ (thin black dashed lines), respectively.
    The adopted parameters are summarized in Table \ref{tbl:parameters}.
    }
    \label{fig:LowSigma}
\end{figure}

Cooling acceleration is important for a relativistically-hot flow.
Fig. \ref{fig:LowSigma} shows the flow profiles for $\sigma_0 = 10^{-2}$ corresponding to $\Theta_0 \approx 2.0 \times 10^3$.
The conversion and dissipation effects are ignored for the thick lines and $x_{\rm cool}$ is 10 (thick red solid lines), 10$^3$ (thick blue dashed lines), and 10$^5$ (thick green dotted lines), respectively.
Only the conversion effect is added for the black dashed line ($x_{\rm conv} = 10^3$), while both $x_{\rm conv}$ and $x_{\rm diss}$ are considered for the black dotted line.
The parameters are summarized in Table \ref{tbl:parameters}.

\subsubsection{Results}\label{sec:CoolingAccelerationResults}

The top panel of Fig. \ref{fig:LowSigma} shows the four velocity profiles.
$u^2_0 \approx 0.5$ is almost the sound velocity of a relativistically-hot fluid.
The flows start to accelerate at $x \approx x_{\rm cool}$ and the terminal velocity of $u_{\infty} \approx 32$ is the same for the different values of $x_{\rm cool}$ (see the thick lines).
The thin black dotted line is almost overlapped with the thick dashed blue line, while $u_{\infty} \approx 13$ for the thin black dashed line is significantly small compared with that for the thick lines.
In the acceleration phase, the four velocity profiles obey $u \propto x^{3/4}$ which is consistent with the analytic estimate introduced in appendix \ref{app:CoolingAcceleration}, i.e., $u \propto x^{1/\hat{\Gamma}}$.

The middle and bottom panels of Fig. \ref{fig:LowSigma} are the magnetization and the total energy flux profiles, respectively ($l_0 = u_0 \gamma_0 \epsilon_0 = 1$).
The terminal values are $\sigma_{\infty} \approx 2$ and $l_{\infty} \approx 10^{-2}$ for all the thick lines, i.e., $\approx$ 99 \% of the jet power is taken away from the system.
The terminal values are higher for the thin black dashed line and lower for the thin black dotted lines than the thick lines.

\subsubsection{Discussion}\label{sec:CoolingAccelerationDiscussion}

All the thick lines overlap with each other when we normalize $x$ with $x_{\rm cool}$ because the characteristic length-scale of the system is only $x_{\rm cool}$ for the thick lines (cf., appendix \ref{app:CoolingAcceleration}).
It means that all the terminal values ($u_{\infty}, \sigma_{\infty}, l_{\infty}$) are exactly the same and depend only on the inlet boundary condition for $x_{\rm conv}, x_{\rm diss} \rightarrow \infty$.
For example of $\sigma_0 = 10^{-3}$, $u_0$, $\Theta_0$ and $u_{\infty}$ are almost the same as those of $\sigma_0 = 10^{-2}$ while both $\sigma_{\infty}$ and $l_{\infty}$ are smaller than those of $\sigma_0 = 10^{-2}$ (cf. Fig. \ref{fig:SgmDependence} in appendix \ref{app:VariousInletBoundary}).
When more than two characteristic length-scales are introduced to the system, the terminal values are different for their combinations even when the inlet boundary is the same (the thin lines in Fig. \ref{fig:LowSigma}).

The flow tends to be cold ($w_{\infty} = \rho_{\infty}$) for a finite $x_{\rm cool}$ and then the relation between the terminal values is written as,
\begin{eqnarray}\label{eq:TerminalValuesCoolingAcceleration}
	\gamma_{\infty} (1 + \sigma_{\infty})
	=
	\gamma_{\rm max} l_{\infty},
\end{eqnarray}
where $l_0 = 1$ and equation (\ref{eq:MassFlux}) are used.
Equation (\ref{eq:TerminalValuesCoolingAcceleration}) just represents the relation between the resultant terminal values.
The terminal values are not obtained without solving the differential equations as follows.

Cooling acceleration is triggered by the outward pressure gradient force which results from extracting the plasma heat by the cooling effect.
Acceleration reduces the density according to equation (\ref{eq:MassFlux}) and also reduces the pressure according to equation (\ref{eq:InternalEnergyConservation}), i.e., acceleration itself is also a source of the outward pressure gradient force.
The flow reaches a terminal velocity when the flow becomes cold $p \rightarrow 0$.
Estimating $u_{\infty}$ is difficult without solving the differential equations because the internal energy density decreases not only by the cooling effect but also by flow acceleration.
Although the inlet and terminal internal energy density $e_{\rm int,0} \approx 0.84$ and $e_{\rm int,\infty} = 0$ are the same for all the lines in Fig. \ref{fig:LowSigma}, the energy taken away from the system by the cooling effect $(l_0 - l_{\infty})$ is clearly different for different for the cases of the thin black lines because the fractions of $e_{\rm int,0}$ converted into the kinetic ($u_{\infty}$) and also into the magnetic energy ($\sigma_{\infty}$) are different.

The conversion effect does not work as an acceleration mechanism, or even it lowers the terminal velocity for the thin black dashed line on the top panel of Fig. \ref{sec:AccelerationMechanisms}.
Because $\sigma_0 \ll 1$ and $x_{\rm conv} = x_{\rm cool}$, cooling acceleration always dominates over conversion acceleration.
The discrepancy of the velocity profile from the thick lines becomes apparent when the magnetization approaches unity and this relates with the different reactions of the toroidal and turbulent magnetic field to flow acceleration.
From the left-hand side of equations (\ref{eq:ToroidalMagneticEnergyConservation}) and (\ref{eq:TurbulentMagneticEnergyConservation}), $\bar{b}^2 u^2$ and $\delta b^2 u^{3/4}$ are constants along the flow if we ignore the right-hand side of these equations.
This represents that $\bar{b}^2$ and $\delta b^2$ behave as the fluid components of their adiabatic indices of 2 (two degrees of freedom gas) and $3/4$ (relativistic photon gas), respectively \citep[cf.][]{Spitzer56}.
In the acceleration phase, $\delta b^2$ decreases slower than $\bar{b}^2$ and then the terminal magnetization $\sigma_{\infty}$ is about an order of magnitude larger for the black dashed line cases than the thick line cases.

\subsection{Conversion acceleration}\label{sec:ConversionAcceleration}

%
\begin{figure}
	\includegraphics[width=\columnwidth]{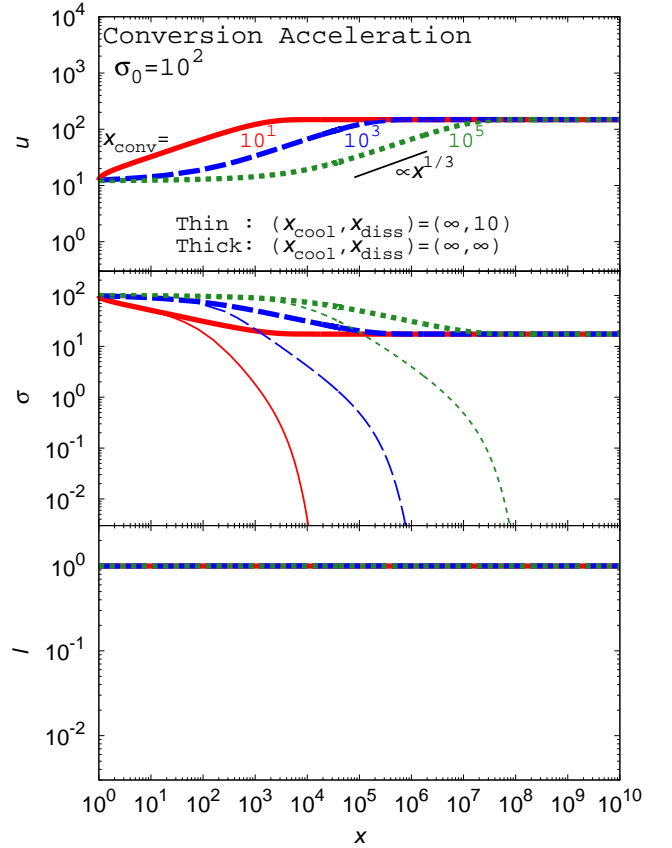}
    \caption{
    The flow profiles when conversion acceleration is dominant.
    The cooling effect is omitted $x_{\rm cool} = \infty$ and is $\sigma_0 = 10^{2}$.
    The results for $x_{\rm conv} =$ 10 (red solid lines), 10$^3$ (blue dashed lines), and 10$^5$ (green dotted lines) are shown.
    The thin lines which are the case of $x_{\rm diss} = 10$ are overplotted on no dissipation case ($x_{\rm diss} = \infty$: thick lines).
    The adopted parameters are summarized in Table \ref{tbl:parameters}.
    }
    \label{fig:Conversion}
\end{figure}

Conversion acceleration is important for the high-$\sigma_0$ case.
Fig. \ref{fig:Conversion} shows the flow profiles for $\sigma_0 = 10^2$ with $x_{\rm conv} =$ 10 (red solid lines), 10$^3$ (blue dashed lines), and 10$^5$ (green dotted lines), respectively.
The cooling effect is omitted for all the lines ($x_{\rm cool} = \infty$).
We consider dissipation of the turbulent magnetic field for the thin lines ($x_{\rm diss} = 10$), while we set no dissipation for the thick lines ($x_{\rm diss} = \infty$).
The adopted parameters and the derived terminal values are summarized in Table \ref{tbl:parameters}.

\subsubsection{Results}\label{sec:ConversionAccelerationResults}

The top panel of Fig. \ref{fig:Conversion} shows the four velocity profiles $u(x)$, where the thin and thick lines are overlapped with each other.
The inlet velocity of $u_0 \approx 12$ is much larger than Fig. \ref{fig:LowSigma}, because $u_{\rm c,0} \sim \sqrt{\sigma_0}$ for $\sigma_0 \gg 1$,.
The flows start to accelerate at $x \approx x_{\rm conv}$ and the terminal velocity of $u_{\infty} \approx 148$ is the same for all the lines.
Although the temperature is not ultra-relativistic ($\Theta_0 \approx 2$ and $\hat{\Gamma} > 4/3$), deceleration by the dissipation effect is less important for the thin lines.
This is because the plasma temperature increases by magnetic dissipation so that dissipation deceleration eventually becomes ineffective.
We find $u \propto x^{1/3}$ in the acceleration phase.
This is consistent with the analytic estimate introduced in appendix \ref{app:ConversionAcceleration}.

The middle and bottom panels of Fig. \ref{fig:Conversion} show the profiles of the magnetization $\sigma(x)$ and the total energy flux $l(x)$.
Difference between the thin and thick lines is evident only in the magnetization profile, and $\sigma_{\infty} \approx 17 \ll \sigma_0$ even without magnetic dissipation for the thick lines.
Because $x_{\rm cool} = \infty$, the total energy flux is conserved $l(x) = l_0 = 1$ for all the lines, i.e., $\epsilon(x)$ decreases with flow acceleration.

\subsubsection{Discussion}\label{sec:ConversionAccelerationDiscussion}

As is the case of cooling acceleration, the system can be normalized by the characteristic length-scale $x_{\rm conv}$ for the thick lines and all the terminal values are the same for the different $x_{\rm conv}$ (see also appendix \ref{app:ConversionAcceleration}).
The terminal values again depend on the inlet boundary condition (cf. Fig. \ref{fig:SgmDependence} in appendix \ref{app:VariousInletBoundary}).
Both of the inlet and terminal velocities are lower for lower $\sigma_0$ and conversion acceleration does not work for $\sigma_0 \ll 1$.

Conservation of the mass and energy fluxes leads to
\begin{eqnarray}\label{eq:TerminalValuesConversionAcceleration}
	\gamma_{\infty} \omega(\Theta_{\infty}) (1 + \sigma_{\infty})
	=
	\gamma_{\rm max}.
\end{eqnarray}
Even for the case of a finite $x_{\rm diss}$ ($\sigma_{\infty} \rightarrow 0$), we do not find an analytical way to estimate the terminal velocity because $\Theta_{\infty} \approx 17$ also increases from the inlet value by magnetic dissipation.

The conversion effect also induces the outward (magnetic) pressure gradient force.
Without magnetic dissipation, it is evident from the second term of the left-hand side of equations (\ref{eq:ToroidalMagneticEnergyConservation}) and (\ref{eq:TurbulentMagneticEnergyConservation}) that the conversion effect does not change the magnetic energy density ($e_{\bar{b}} + e_{\delta b} = \bar{b}^2/2 + \delta b^2/2$) but changes the `magnetic pressure' ($p_{\bar{b}} + p_{\delta b} = \bar{b}^2/2+\delta b^2/6$).
This also results from the different effective adiabatic indices between the toroidal and turbulent magnetic field as discussed in section \ref{sec:CoolingAccelerationDiscussion}.
Here, the enthalpies for the toroidal and turbulent magnetic field are $e_{\bar{b}} + p_{\bar{b}} = \bar{b}^2$ and $e_{\delta b} + p_{\delta b} = (2/3)\delta b^2$ (see equation (\ref{eq:TotalEnergyDensity})).
As a result, the outward pressure gradient accelerates the flow until all the toroidal magnetic field is totally converted into the turbulent one.

\subsection{Efficient acceleration}\label{sec:EfficientAcceleration}

%
\begin{figure}
	\includegraphics[width=\columnwidth]{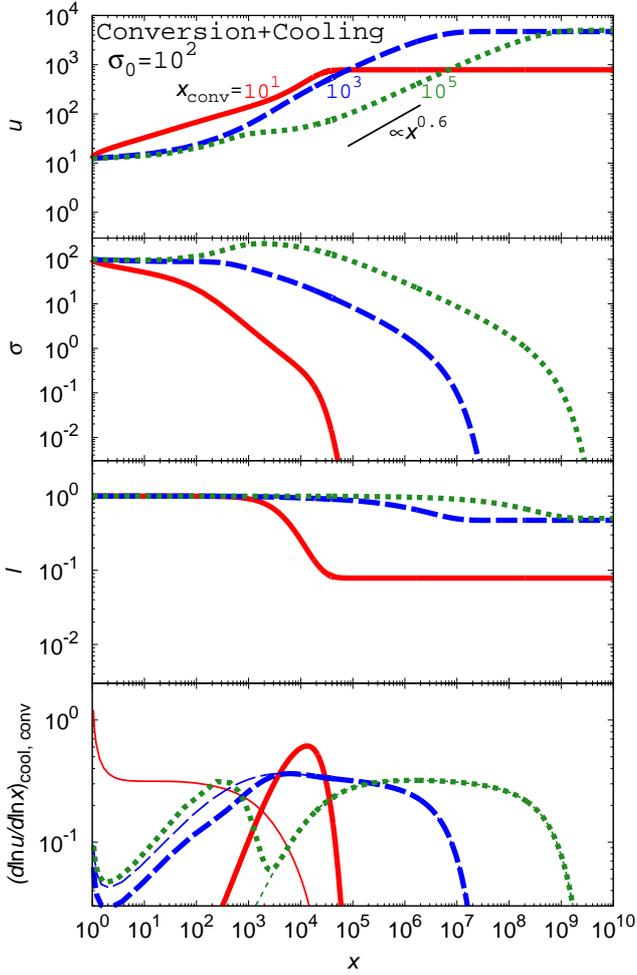}
    \caption{
    The flow profiles for efficiently accelerated cases, where all the cooling, conversion and dissipation effects play a role.
    The flows are high-$\sigma_0$ ($\sigma_0 = 10^{2}$).
    The results for $x_{\rm conv} =$ 10 (red solid lines), 10$^3$ (blue dashed lines), and 10$^5$ (green dotted lines) are shown.
    $x_{\rm diss} = x_{\rm cool} = 10$ are common for all the lines.
    The adopted parameters are summarized in Table \ref{tbl:parameters}.
    The thin and thick lines in the bottom panel correspond to the conversion and cooling acceleration rates, respectively (see equation (\ref{eq:CoolingAccelerationRate})).
    }
    \label{fig:Both}
\end{figure}

The high-$\sigma_0$ flows which experienced conversion acceleration can be accelerated by the cooling effect after the dissipation of the turbulent magnetic field into the plasma heat.
Fig. \ref{fig:Both} shows the high-$\sigma_0$ flow profiles with both of the conversion and cooling effects.
$x_{\rm conv} =$ 10 (red solid), 10$^3$ (blue dashed), and 10$^5$ (green dotted) are different for each line, while $x_{\rm diss} = x_{\rm cool} = 10$ are common for all the lines.
The parameters of the calculations are summarized in Table \ref{tbl:parameters}.
Note that both cooling and conversion acceleration work even for the other combinations of $(x_{\rm cool}, x_{\rm conv}, x_{\rm diss})$ but the flows are not always accelerated efficiently as Fig. \ref{fig:Both}.

\subsubsection{Results}\label{sec:EfficientAccelerationResults}

The top panel of Fig. \ref{fig:Both} shows the four velocity profiles $u(x)$.
Additional acceleration by cooling increases the terminal velocity compared with Fig. \ref{fig:Conversion}.
For the red solid line ($x_{\rm conv} = 10$), in addition to conversion acceleration at $x \lesssim 10^3$, cooling acceleration at $x \sim 10^4$ increases the terminal velocity to $u_{\infty} \approx 786$ which is about five times larger than that of conversion acceleration.
For the blue dashed and green dotted lines ($x_{\rm conv} = 10^3,~10^5$), their terminal velocities attain to about half of the maximum possible value of $u_{\infty} \approx \gamma_{\rm max} / 2$.
The acceleration profiles for the blue and green lines are different from both cooling $u \propto x^{1/\hat{\Gamma}}$ and conversion $\propto x^{1/3}$ acceleration, and are close to $u \propto x^{0.6}$ (see below).

The second and third panels of Fig. \ref{fig:Both} show the profiles of the magnetization $\sigma(x)$ and the total energy flux $l(x)$.
For the green dotted line in the second panel, the magnetization increases slightly at $x \lesssim x_{\rm conv} = 10^5$ because of the cooling effect.
The terminal value of the total energy flux is much larger than the case of cooling acceleration (Fig. \ref{fig:LowSigma}) and $l_{\infty} \approx 0.5$ for the blue and green lines.
The radiation efficiency is about 50\%.

The bottom panel of Fig. \ref{fig:Both} shows the flow acceleration rates by the conversion (thin) and cooling (thick) effects, respectively.
For example, the flow acceleration rate by the cooling effect is written as (cf. equations (\ref{eq:DifferentialEquationForVelocity}) and (\ref{eq:CoolingEffect}))
\begin{eqnarray}\label{eq:CoolingAccelerationRate}
	\left( \frac{d \ln u}{d \ln x} \right)_{\rm cool}
	=
	\frac{(\hat{\Gamma} - 1) e_{\rm int}}{u \epsilon (\beta^2 - \beta^2_c)} \frac{x}{x_{\rm cool}}.
\end{eqnarray}
The flow deceleration rate by the dissipation effect is not plotted for clarity and is always subdominant.
At the first acceleration phase $x \lesssim 10^3$, conversion acceleration plays a role for the red solid and blue dashed lines while cooling acceleration works for the green dotted line.
Comparing the red solid with the blue dashed and green dotted lines, we find that the second acceleration phase ($x \gtrsim 10^3$) is important for efficient acceleration, where both of cooling and conversion acceleration equally work.
The combined flow acceleration rate becomes $(d \ln u / d \ln x)_{\rm cool + conv} \approx 0.6$ for the second acceleration phase of the blue dashed and green dotted lines (see also the top panel of Fig. \ref{fig:Both}).

\subsubsection{Discussion}\label{sec:EfficientAccelerationDiscussion}

Including all the non-ideal MHD effects, the flow tends to be cold ($w_{\infty} = \rho_{\infty}$) and unmagnetized ($\sigma_{\infty} = 0$).
The terminal Lorentz factor and the terminal total energy flux have a simple relation
\begin{eqnarray}\label{eq:TerminalValuesEfficientAcceleration}
	\gamma_{\infty}
	=
	\gamma_{\rm max} l_{\infty}.
\end{eqnarray}
Nevertheless, we find the terminal velocity only by numerically solving the differential equations.
The dependence on the inlet boundary condition are studied in appendix \ref{app:VariousInletBoundary} (Fig. \ref{fig:SgmDependence}).
Although the terminal velocity decreases with $\sigma_0$, it can be a still significant fraction of $\gamma_{\rm max}$ ($u_{\infty} \approx \gamma_{\rm max} / 5$) for $\sigma_0 = 1$.

Some of our results are similar to those of \citet{Drenkhahn&Spruit02}, i.e., the terminal Lorentz factor is about half of the maximum possible value and about half of the inlet total energy flux is radiated away from the system for their high-$\sigma$ cases (their Fig. 3).
Our efficient acceleration is similar to the calculations of \citet{Drenkhahn&Spruit02} when $x_{\rm conv}$ is much larger than both $x_{\rm cool}$ and $x_{\rm diss}$ because \citet{Drenkhahn&Spruit02} set fast cooling and direct dissipation of the toroidal magnetic field into the plasma heat.
On the other hand, our acceleration rate ($u \propto x^{0.6}$) is faster than theirs that is slower than $u \propto r^{1/3}$ (the top-panel of their Fig. 1).
Differences between our and their calculations are that they adopted spherical geometry and the dissipation time-scale is not constant in the proper frame like our formulation but is almost constant in the observer frame.

For spherical geometry, the expansion acceleration term appears in equation (\ref{eq:DifferentialEquationForVelocity}) \citep[see equation (19) of][]{Tanaka+18}.
The analytic result $u \propto r^{1/3}$ by \citet{Drenkhahn02} is reproduced without both cooling and expansion acceleration while we should assume that the spherical flow is always cold because of adiabatic expansion (see appendix \ref{app:DissipationAcceleration}).
The flow also remains cold for \citet{Drenkhahn&Spruit02} because of fast cooling similar to the present case, and then the difference between our acceleration rate and theirs might not be caused by difference of adopted geometry in this sense.
More systematic studies are required to determine the acceleration mechanism of \citet{Drenkhahn&Spruit02}.


\section{Discussion and Conclusions}\label{sec:dis_cons}

Acceleration of cylindrical jets is studied by including the three non-ideal MHD effects, namely the cooling $x_{\rm cool}$, conversion $x_{\rm conv}$, and dissipation effects $x_{\rm diss}$ \citep[cf.][]{Tanaka+18}.
For a Synge gas which has the effective adiabatic index of $4/3 \le \hat{\Gamma} \le 5/3$, the cooling and conversion effects accelerate the flow of $\beta > \beta_{\rm c}$ while the dissipation effect decelerates it.
Although each of cooling and conversion acceleration does not accelerate the flow significantly (Figs. \ref{fig:LowSigma} and \ref{fig:Conversion}), the terminal velocity attains about half of the maximum possible value of $\gamma_{\rm max} / 2$ when all the three effects work simultaneously (Fig. \ref{fig:Both}).

The model of the broadband emission of blazers implies that the jets are particle dominated flows and the radiative efficiency of them would be a few tens of percent of their kinetic luminosity \citep[e.g.,][]{Celotti&Ghisellini08}.
Based on the ideal MHD picture of relativistic jets, shock dissipation of the kinetic energy of the particle dominated jet results in the emission, e.g., the reconfinement shock model \citep[][]{Sanders83}.
Efficient radiation means significant deceleration of the jet by the shock \citep[e.g,][]{Nalewajko12}.
Our non-ideal MHD picture for the efficiently accelerated case is more like magnetic reconnection rather than the shock as the dissipation process.
Dissipation and subsequent cooling processes which result from an internal instability of the jet \citep[e.g.,][]{Bromberg&Tchekhovskoy16} accelerate the flow.
The radiative efficiency will be about a few tens percent of the jet luminosity for our efficiently accelerated case.

The blazer emission also implies that the emission region is relatively close to the supermassive black-hole ($10^3 - 10^4 r_{\rm g}$) \citep[e.g.,][]{Chiaberge&Ghisellini99}.
Although \citet{Drenkhahn&Spruit02} also found the solutions which has similar terminal velocity $\sim \gamma_{\rm max} / 2$ and similar radiative efficiency $\sim l_0 / 2$ to ours in the spherical outflow, our acceleration rate $\propto x^{0.6}$ is faster than theirs that is slower than $\propto r^{1/3}$.
In addition, the trans-characteristic-velocity point $z_0$ in cylindrical geometry does not relate directly with and could be smaller than the Schwarzschild radius $r_{\rm g}$ because the origin of coordinates need not be the center of the supermassive black hole.
This is not the case in spherical geometry in which the trans-characteristic-velocity `radius' should be larger than $r_{\rm g}$.


In the present paper, we adopt the cylindrical geometry for simplicity and also for the purpose of studying the non-ideal MHD acceleration.
For the other jet geometries, the non-ideal MHD acceleration also works in addition to the ordinary `expansion acceleration' in ideal MHD.
In practice, we should consider the lateral jet structure by solving the transfield force balance equation \citep[e.g.,][]{Lyubarsky11} and look for the relative importance of the ideal to non-ideal MHD acceleration, simultaneously.
Generalization of the transfield force balance equation by including the phenomenological non-ideal MHD effects are beyond the scope of the present study and the generalization is expected based on more sophisticated formulation of non-ideal MHD effects. 

However, it is still worth discussing the lateral force balance in cylindrical jets.
In order to keep the cylindrical structure, the lateral force balance equation (radial component of equation (\ref{eq:app:EnergyMomentumConservationCooling})) should be satisfied all along the jet axis, i.e.,
\begin{eqnarray}\label{eq:TransverseForceBalance}
	\nabla_{\mu} T^{\mu R}
	& = &
	\frac{d}{d R} \left(p + \frac{\delta b^2}{6} \right) + \frac{\bar{b}}{R} \frac{d}{d R} R \bar{b} \nonumber \\
	& = &
	\frac{d}{d R} \left(p + \frac{\delta b^2}{6} + \frac{\bar{b}^2}{2} \right) + \frac{\bar{b}^2}{R}
	= 0,
\end{eqnarray}
for any $z$.
The toroidal magnetic field has clearly different contribution because of the magnetic tension force.
We first consider the case that the transverse force balance is satisfied at the inlet boundary and that the external pressure $p_{\rm ext}$ is constant along the jet axis.
The ideal MHD flow can be cylindrical because all $p,~\delta b^2,~\bar{b}^2$ are independent from $z$.
The cooling term alone reduces $p$ along the flow, i.e., we have a converging flow.
The dissipation term alone converts $\delta b^2 / 2$ to $e_{\rm int} \approx 3 p$ (for relativistically hot plasma) and then the flow can keep an approximately cylindrical structure.
The conversion term alone converts $\bar{b}^2 / 2$ to $\delta b^2 / 2$ and then we have a diverging flow by the reduction of the magnetic tension.
Efficiently accelerating, approximately cylindrical jets would be realized in special situations, such as the mildly magnetized flow $\sigma_0 \le 1$ requiring the magnetic tension force to be subdominant with decreasing $p_{\rm ext}$ against extraction of the internal pressure by radiation along the jet.

%

Our non-ideal MHD formulation is heuristic but would be a way to fill the gap between simplified analytic (ideal MHD) studies and more realistic numerical studies.
There are some other non-ideal MHD terms which make both analytical and numerical (theoretical) studies close to the reality.
Mass loading to the jet is an example \cite[cf.][]{Komissarov94, Toma&Takahara12}.
Effects of multi-composition plasma would also be significant to change the dynamics of relativistic jets.
For example, the three-fluid (electron, proton and charged dust) model by \citet{Motschmann+92} would provide some insight into electron-positron-proton system.

Effects of cosmic-rays would also be important because they are also considered as an additional component in the MHD formulation \citep[cf.][]{Bai+15}.
The cooling effect in this study may be recognized as not only the radiative cooling but also the cosmic-rays (non-thermal particles) production and escape because the energy of the thermal particles is reduced by the back-reaction of cosmic-ray acceleration.
Before we observe the (non-thermal) radiative signatures from the relativistic jets, there must occur particle acceleration and it should acquire a non-negligible fraction of the jet luminosity.
Stochastic particle acceleration by the magnetic turbulence is expected and is a promising process to explain the non-thermal emission from some relativistic objects including blazers \citep[e.g,][]{Asano+14, Sasaki&Asano15, Tanaka&Asano17}.




\section*{Acknowledgements}

S. J. T. would like to thank K. Asano, R. Yamazaki, K. Kisaka and T. Terasawa for useful discussion.
This work is supported by JSPS Grants-in-Aid for Scientific Research Nos. 17H18270 (SJT) and 18H01245 (KT).



\bibliographystyle{mnras}
\bibliography{draft} 

\begin{thebibliography}{}
\makeatletter
\relax
\def\mn@urlcharsother{\let\do\@makeother \do\$\do\&\do\#\do\^\do\_\do\%\do\~}
\def\mn@doi{\begingroup\mn@urlcharsother \@ifnextchar [ {\mn@doi@}
  {\mn@doi@[]}}
\def\mn@doi@[#1]#2{\def\@tempa{#1}\ifx\@tempa\@empty \href
  {http://dx.doi.org/#2} {doi:#2}\else \href {http://dx.doi.org/#2} {#1}\fi
  \endgroup}
\def\mn@eprint#1#2{\mn@eprint@#1:#2::\@nil}
\def\mn@eprint@arXiv#1{\href {http://arxiv.org/abs/#1} {{\tt arXiv:#1}}}
\def\mn@eprint@dblp#1{\href {http://dblp.uni-trier.de/rec/bibtex/#1.xml}
  {dblp:#1}}
\def\mn@eprint@#1:#2:#3:#4\@nil{\def\@tempa {#1}\def\@tempb {#2}\def\@tempc
  {#3}\ifx \@tempc \@empty \let \@tempc \@tempb \let \@tempb \@tempa \fi \ifx
  \@tempb \@empty \def\@tempb {arXiv}\fi \@ifundefined
  {mn@eprint@\@tempb}{\@tempb:\@tempc}{\expandafter \expandafter \csname
  mn@eprint@\@tempb\endcsname \expandafter{\@tempc}}}

\bibitem[\protect\citeauthoryear{{Asada} \& {Nakamura}}{{Asada} \&
  {Nakamura}}{2012}]{Asada&Nakamura12}
{Asada} K.,  {Nakamura} M.,  2012, \mn@doi [\apjl]
  {10.1088/2041-8205/745/2/L28}, \href
  {https://ui.adsabs.harvard.edu/abs/2012ApJ...745L..28A} {745, L28}

\bibitem[\protect\citeauthoryear{{Asada}, {Nakamura}  \& {Pu}}{{Asada}
  et~al.}{2016}]{Asada+16}
{Asada} K.,  {Nakamura} M.,   {Pu} H.-Y.,  2016, \mn@doi [\apj]
  {10.3847/1538-4357/833/1/56}, \href
  {https://ui.adsabs.harvard.edu/abs/2016ApJ...833...56A} {833, 56}

\bibitem[\protect\citeauthoryear{{Asano}, {Takahara}, {Kusunose}, {Toma}  \&
  {Kakuwa}}{{Asano} et~al.}{2014}]{Asano+14}
{Asano} K.,  {Takahara} F.,  {Kusunose} M.,  {Toma} K.,   {Kakuwa} J.,  2014,
  \mn@doi [\apj] {10.1088/0004-637X/780/1/64}, \href
  {https://ui.adsabs.harvard.edu/abs/2014ApJ...780...64A} {780, 64}

\bibitem[\protect\citeauthoryear{{Bai}, {Caprioli}, {Sironi}  \&
  {Spitkovsky}}{{Bai} et~al.}{2015}]{Bai+15}
{Bai} X.-N.,  {Caprioli} D.,  {Sironi} L.,   {Spitkovsky} A.,  2015, \mn@doi
  [\apj] {10.1088/0004-637X/809/1/55}, \href
  {https://ui.adsabs.harvard.edu/abs/2015ApJ...809...55B} {809, 55}

\bibitem[\protect\citeauthoryear{{Beskin}}{{Beskin}}{2010}]{Beskin10}
{Beskin} V.~S.,  2010, {MHD Flows in Compact Astrophysical Objects}.
Astronomy and Astrophysics Library, Springer, Berlin,
  \mn@doi{10.1007/978-3-642-01290-7}

\bibitem[\protect\citeauthoryear{{Beskin} \& {Nokhrina}}{{Beskin} \&
  {Nokhrina}}{2009}]{Beskin&Nokhrina09}
{Beskin} V.~S.,  {Nokhrina} E.~E.,  2009, \mn@doi [\mnras]
  {10.1111/j.1365-2966.2009.14964.x}, \href
  {https://ui.adsabs.harvard.edu/abs/2009MNRAS.397.1486B} {397, 1486}

\bibitem[\protect\citeauthoryear{{Bromberg} \& {Tchekhovskoy}}{{Bromberg} \&
  {Tchekhovskoy}}{2016}]{Bromberg&Tchekhovskoy16}
{Bromberg} O.,  {Tchekhovskoy} A.,  2016, \mn@doi [\mnras]
  {10.1093/mnras/stv2591}, \href
  {https://ui.adsabs.harvard.edu/abs/2016MNRAS.456.1739B} {456, 1739}

\bibitem[\protect\citeauthoryear{{Bromberg}, {Singh}, {Davelaar}  \&
  {Philippov}}{{Bromberg} et~al.}{2019}]{Bromberg+19}
{Bromberg} O.,  {Singh} C.~B.,  {Davelaar} J.,   {Philippov} A.~A.,  2019,
  arXiv e-prints, \href {https://ui.adsabs.harvard.edu/abs/2019arXiv190808620B}
  {p. arXiv:1908.08620}

\bibitem[\protect\citeauthoryear{{Celotti} \& {Ghisellini}}{{Celotti} \&
  {Ghisellini}}{2008}]{Celotti&Ghisellini08}
{Celotti} A.,  {Ghisellini} G.,  2008, \mn@doi [\mnras]
  {10.1111/j.1365-2966.2007.12758.x}, \href
  {https://ui.adsabs.harvard.edu/abs/2008MNRAS.385..283C} {385, 283}

\bibitem[\protect\citeauthoryear{{Chiaberge} \& {Ghisellini}}{{Chiaberge} \&
  {Ghisellini}}{1999}]{Chiaberge&Ghisellini99}
{Chiaberge} M.,  {Ghisellini} G.,  1999, \mn@doi [\mnras]
  {10.1046/j.1365-8711.1999.02538.x}, \href
  {https://ui.adsabs.harvard.edu/abs/1999MNRAS.306..551C} {306, 551}

\bibitem[\protect\citeauthoryear{{Coroniti}}{{Coroniti}}{1990}]{Coroniti90}
{Coroniti} F.~V.,  1990, \mn@doi [\apj] {10.1086/168340}, \href
  {https://ui.adsabs.harvard.edu/abs/1990ApJ...349..538C} {349, 538}

\bibitem[\protect\citeauthoryear{{Drenkhahn}}{{Drenkhahn}}{2002}]{Drenkhahn02}
{Drenkhahn} G.,  2002, \mn@doi [\aap] {10.1051/0004-6361:20020390}, \href
  {https://ui.adsabs.harvard.edu/abs/2002A&A...387..714D} {387, 714}

\bibitem[\protect\citeauthoryear{{Drenkhahn} \& {Spruit}}{{Drenkhahn} \&
  {Spruit}}{2002}]{Drenkhahn&Spruit02}
{Drenkhahn} G.,  {Spruit} H.~C.,  2002, \mn@doi [\aap]
  {10.1051/0004-6361:20020839}, \href
  {https://ui.adsabs.harvard.edu/abs/2002A&A...391.1141D} {391, 1141}

\bibitem[\protect\citeauthoryear{{Ghisellini}, {Tavecchio}, {Foschini},
  {Ghirland a}, {Maraschi}  \& {Celotti}}{{Ghisellini}
  et~al.}{2010}]{Ghisellini+10}
{Ghisellini} G.,  {Tavecchio} F.,  {Foschini} L.,  {Ghirland a} G.,  {Maraschi}
  L.,   {Celotti} A.,  2010, \mn@doi [\mnras]
  {10.1111/j.1365-2966.2009.15898.x}, \href
  {https://ui.adsabs.harvard.edu/abs/2010MNRAS.402..497G} {402, 497}

\bibitem[\protect\citeauthoryear{{Hada}}{{Hada}}{2017}]{Hada17}
{Hada} K.,  2017, \mn@doi [Galaxies] {10.3390/galaxies5010002}, \href
  {https://ui.adsabs.harvard.edu/abs/2017Galax...5....2H} {5, 2}

\bibitem[\protect\citeauthoryear{{Hada} et~al.,}{{Hada} et~al.}{2013}]{Hada+13}
{Hada} K.,  et~al., 2013, \mn@doi [\apj] {10.1088/0004-637X/775/1/70}, \href
  {https://ui.adsabs.harvard.edu/abs/2013ApJ...775...70H} {775, 70}

\bibitem[\protect\citeauthoryear{{Inoue} \& {Takahara}}{{Inoue} \&
  {Takahara}}{1996}]{Inoue&Takahara96}
{Inoue} S.,  {Takahara} F.,  1996, \mn@doi [\apj] {10.1086/177270}, \href
  {https://ui.adsabs.harvard.edu/abs/1996ApJ...463..555I} {463, 555}

\bibitem[\protect\citeauthoryear{{J{\"u}ttner}}{{J{\"u}ttner}}{1928}]{Juttner28}
{J{\"u}ttner} F.,  1928, \mn@doi [Zeitschrift fur Physik] {10.1007/BF01340339},
  \href {https://ui.adsabs.harvard.edu/abs/1928ZPhy...47..542J} {47, 542}

\bibitem[\protect\citeauthoryear{{Komissarov}}{{Komissarov}}{1994}]{Komissarov94}
{Komissarov} S.~S.,  1994, \mn@doi [\mnras] {10.1093/mnras/269.2.394}, \href
  {https://ui.adsabs.harvard.edu/abs/1994MNRAS.269..394K} {269, 394}

\bibitem[\protect\citeauthoryear{{Komissarov}, {Barkov}, {Vlahakis}  \&
  {K{\"o}nigl}}{{Komissarov} et~al.}{2007}]{Komissarov+07}
{Komissarov} S.~S.,  {Barkov} M.~V.,  {Vlahakis} N.,   {K{\"o}nigl} A.,  2007,
  \mn@doi [\mnras] {10.1111/j.1365-2966.2007.12050.x}, \href
  {https://ui.adsabs.harvard.edu/abs/2007MNRAS.380...51K} {380, 51}

\bibitem[\protect\citeauthoryear{{Komissarov}, {Vlahakis}, {K{\"o}nigl}  \&
  {Barkov}}{{Komissarov} et~al.}{2009}]{Komissarov+09}
{Komissarov} S.~S.,  {Vlahakis} N.,  {K{\"o}nigl} A.,   {Barkov} M.~V.,  2009,
  \mn@doi [\mnras] {10.1111/j.1365-2966.2009.14410.x}, \href
  {https://ui.adsabs.harvard.edu/abs/2009MNRAS.394.1182K} {394, 1182}

\bibitem[\protect\citeauthoryear{{Lyubarsky}}{{Lyubarsky}}{2009}]{Lyubarsky09}
{Lyubarsky} Y.,  2009, \mn@doi [\apj] {10.1088/0004-637X/698/2/1570}, \href
  {https://ui.adsabs.harvard.edu/abs/2009ApJ...698.1570L} {698, 1570}

\bibitem[\protect\citeauthoryear{{Lyubarsky}}{{Lyubarsky}}{2011}]{Lyubarsky11}
{Lyubarsky} Y.,  2011, \mn@doi [\pre] {10.1103/PhysRevE.83.016302}, \href
  {https://ui.adsabs.harvard.edu/abs/2011PhRvE..83a6302L} {83, 016302}

\bibitem[\protect\citeauthoryear{{Lyubarsky} \& {Kirk}}{{Lyubarsky} \&
  {Kirk}}{2001}]{Lyubarsky&Kirk01}
{Lyubarsky} Y.,  {Kirk} J.~G.,  2001, \mn@doi [\apj] {10.1086/318354}, \href
  {https://ui.adsabs.harvard.edu/abs/2001ApJ...547..437L} {547, 437}

\bibitem[\protect\citeauthoryear{{Motschmann}, {Sauer}  \&
  {Roatsch}}{{Motschmann} et~al.}{1992}]{Motschmann+92}
{Motschmann} U.,  {Sauer} K.,   {Roatsch} T.,  1992, \mn@doi [\grl]
  {10.1029/91GL03166}, \href
  {https://ui.adsabs.harvard.edu/abs/1992GeoRL..19..225M} {19, 225}

\bibitem[\protect\citeauthoryear{{Nakamura} et~al.,}{{Nakamura}
  et~al.}{2018}]{Nakamura+18}
{Nakamura} M.,  et~al., 2018, \mn@doi [\apj] {10.3847/1538-4357/aaeb2d}, \href
  {https://ui.adsabs.harvard.edu/abs/2018ApJ...868..146N} {868, 146}

\bibitem[\protect\citeauthoryear{{Nalewajko}}{{Nalewajko}}{2012}]{Nalewajko12}
{Nalewajko} K.,  2012, \mn@doi [\mnras] {10.1111/j.1745-3933.2011.01193.x},
  \href {https://ui.adsabs.harvard.edu/abs/2012MNRAS.420L..48N} {420, L48}

\bibitem[\protect\citeauthoryear{{Ogihara}, {Takahashi}  \& {Toma}}{{Ogihara}
  et~al.}{2019}]{Ogihara+19}
{Ogihara} T.,  {Takahashi} K.,   {Toma} K.,  2019, \mn@doi [\apj]
  {10.3847/1538-4357/ab1909}, \href
  {https://ui.adsabs.harvard.edu/abs/2019ApJ...877...19O} {877, 19}

\bibitem[\protect\citeauthoryear{{Porth} \& {Komissarov}}{{Porth} \&
  {Komissarov}}{2015}]{Porth&Komissarov15}
{Porth} O.,  {Komissarov} S.~S.,  2015, \mn@doi [\mnras]
  {10.1093/mnras/stv1295}, \href
  {https://ui.adsabs.harvard.edu/abs/2015MNRAS.452.1089P} {452, 1089}

\bibitem[\protect\citeauthoryear{{Sanders}}{{Sanders}}{1983}]{Sanders83}
{Sanders} R.~H.,  1983, \mn@doi [\apj] {10.1086/160760}, \href
  {https://ui.adsabs.harvard.edu/abs/1983ApJ...266...73S} {266, 73}

\bibitem[\protect\citeauthoryear{{Sasaki}, {Asano}  \& {Terasawa}}{{Sasaki}
  et~al.}{2015}]{Sasaki&Asano15}
{Sasaki} K.,  {Asano} K.,   {Terasawa} T.,  2015, \mn@doi [\apj]
  {10.1088/0004-637X/814/2/93}, \href
  {https://ui.adsabs.harvard.edu/abs/2015ApJ...814...93S} {814, 93}

\bibitem[\protect\citeauthoryear{{Sob'yanin}}{{Sob'yanin}}{2017}]{Sobyanin17}
{Sob'yanin} D.~N.,  2017, \mn@doi [\mnras] {10.1093/mnras/stx1767}, \href
  {https://ui.adsabs.harvard.edu/abs/2017MNRAS.471.4121S} {471, 4121}

\bibitem[\protect\citeauthoryear{{Spitzer}}{{Spitzer}}{1956}]{Spitzer56}
{Spitzer} L.,  1956, {Physics of Fully Ionized Gases}.
Interscience, New York, NY

\bibitem[\protect\citeauthoryear{{Synge}}{{Synge}}{1957}]{Synge57}
{Synge} J.~L.,  1957, {The Relativistic Gas}.
North-Holland

\bibitem[\protect\citeauthoryear{{Tanaka} \& {Asano}}{{Tanaka} \&
  {Asano}}{2017}]{Tanaka&Asano17}
{Tanaka} S.~J.,  {Asano} K.,  2017, \mn@doi [\apj] {10.3847/1538-4357/aa6f13},
  \href {https://ui.adsabs.harvard.edu/abs/2017ApJ...841...78T} {841, 78}

\bibitem[\protect\citeauthoryear{{Tanaka}, {Toma}  \& {Tominaga}}{{Tanaka}
  et~al.}{2018}]{Tanaka+18}
{Tanaka} S.~J.,  {Toma} K.,   {Tominaga} N.,  2018, \mn@doi [\mnras]
  {10.1093/mnras/sty1356}, \href
  {https://ui.adsabs.harvard.edu/abs/2018MNRAS.478.4622T} {478, 4622}

\bibitem[\protect\citeauthoryear{{Toma} \& {Takahara}}{{Toma} \&
  {Takahara}}{2012}]{Toma&Takahara12}
{Toma} K.,  {Takahara} F.,  2012, \mn@doi [\apj] {10.1088/0004-637X/754/2/148},
  \href {https://ui.adsabs.harvard.edu/abs/2012ApJ...754..148T} {754, 148}

\bibitem[\protect\citeauthoryear{{Toma} \& {Takahara}}{{Toma} \&
  {Takahara}}{2013}]{Toma&Takahara13}
{Toma} K.,  {Takahara} F.,  2013, \mn@doi [Progress of Theoretical and
  Experimental Physics] {10.1093/ptep/ptt058}, \href
  {https://ui.adsabs.harvard.edu/abs/2013PTEP.2013h3E02T} {2013, 083E02}

\bibitem[\protect\citeauthoryear{{Walker}, {Hardee}, {Davies}, {Ly}  \&
  {Junor}}{{Walker} et~al.}{2018}]{Walker+18}
{Walker} R.~C.,  {Hardee} P.~E.,  {Davies} F.~B.,  {Ly} C.,   {Junor} W.,
  2018, \mn@doi [\apj] {10.3847/1538-4357/aaafcc}, \href
  {https://ui.adsabs.harvard.edu/abs/2018ApJ...855..128W} {855, 128}

\bibitem[\protect\citeauthoryear{{Zhou} \& {Matthaeus}}{{Zhou} \&
  {Matthaeus}}{1990}]{Zhou&Matthaeus90c}
{Zhou} Y.,  {Matthaeus} W.~H.,  1990, \mn@doi [\jgr] {10.1029/JA095iA09p14881},
  \href {https://ui.adsabs.harvard.edu/abs/1990JGR....9514881Z} {95, 14881}

\bibitem[\protect\citeauthoryear{{Zrake} \& {Arons}}{{Zrake} \&
  {Arons}}{2017}]{Zrake&Arons17}
{Zrake} J.,  {Arons} J.,  2017, \mn@doi [\apj] {10.3847/1538-4357/aa826d},
  \href {https://ui.adsabs.harvard.edu/abs/2017ApJ...847...57Z} {847, 57}

\makeatother
\end{thebibliography}



\appendix

\section{A Model of The Cooling, Conversion and Dissipation Effects}\label{app:ModelOfNonIdealMHDEffects} 

We follow the relativistic MHD formulation including the three phenomenological terms introduced in \citet{Tanaka+18}.
Stating from the ideal MHD equations (section \ref{app:IdealMHD}), we summarise their non-ideal MHD formulation in the covariant form (sections \ref{app:RaditiveCooling} $-$ \ref{app:MagneticConversion}).
The derivation of equations (\ref{eq:TotalEnergyConservation}) $-$ (\ref{eq:TurbulentMagneticEnergyConservation}) are summarized in section \ref{app:Derivation}.

\subsection{Ideal MHD equations}\label{app:IdealMHD}

On the ideal MHD approximation, the energy-momentum tensor has a fluid $T^{\mu \nu}_{\rm FL}$ and an electromagnetic $T^{\mu \nu}_{\rm EM}$ parts, i.e.,
\begin{eqnarray}\label{eq:app:EnergyMomentumTensor}
	T^{\mu \nu}
	\equiv
	T^{\mu \nu}_{\rm FL} + T^{\mu \nu}_{\rm EM}
	\equiv
	\left(w + b^2 \right) u^{\mu} u^{\nu} + \left( p + b^2 / 2 \right) g^{\mu \nu} - b^{\mu} b^{\nu},
\end{eqnarray}
where $w$, $p$ and $u^{\mu} = \gamma (1, {\bm \beta})$ are the proper enthalpy density, the pressure and a four velocity of the fluid.
The magnetic field four-vector is written as $b^{\mu} = {}^{\ast}F^{\mu \nu} u_{\nu}$ (${}^{\ast}F^{\mu \nu} = (1/2) e^{\mu \nu \alpha \beta} F_{\alpha \beta}$) with the use of the Levi-Civita tensor $e_{\mu \nu \alpha \beta}$ and the electromagnetic field $F^{\mu \nu}$ tensor.

An ideal MHD system is fully described by equation of state, the conservation of the particle number, the conservations of the energy and momentum, and the induction equation, i.e.,
\begin{eqnarray}
	w
	& = &
	w(p, \rho)
	, \label{eq:app:EOS} \\
	\nabla_{\mu} (\rho u^{\mu})
	& = &
	0
	, \label{eq:app:NumberConservation} \\
	\nabla_{\nu} T^{\mu \nu}
	& = &
	0
	, \label{eq:app:EnergyMomentumConservation} \\
	\nabla_{\nu} {}^{\ast}F^{\mu \nu}
	=
	\nabla_{\nu}
	(b^{\mu} u^{\nu} - b^{\nu} u^{\mu})
	& = &
	0
	, \label{eq:app:InductionEquation}
\end{eqnarray}
where $\rho$ is the the proper mass density.

\subsection{Raditive Cooling}\label{app:RaditiveCooling}

The energy-momentum of the MHD system is not preserved by including radiative cooling.
Introducing the cooling rate $\Lambda_{\rm cool}$, we rewrite equation (\ref{eq:app:EnergyMomentumConservation}) as \citep[cf.][]{Drenkhahn&Spruit02}
\begin{eqnarray}
	\nabla_{\nu} T^{\mu \nu}
	& = &
	- u^{\mu} \frac{\Lambda_{\rm cool}}{c}
	, \label{eq:app:EnergyMomentumConservationCooling}
\end{eqnarray}
where $\Lambda_{\rm cool}$ is a parameter of the system and the radiation is isotropic in the comoving frame in this form.

\subsection{Magnetic Dissipation}\label{app:MagneticDissipation}

We adopt a phenomenological formulation of magnetic dissipation (a sink term in the induction equation) by \citet{Drenkhahn02}.
The magnetic field is separated into the non-reconnecting $\hat{b}^{\mu}$ and the decayable $\check{b}^{\mu}$ parts ($b^{\mu} = \hat{b}^{\mu} + \check{b}^{\mu}$), considering the magnetic reconnection of the alternating toroidal magnetic field in the striped wind model \citep[][]{Coroniti90}.
Magnetic dissipation appears as a sink term in the induction equation for $\check{b}^{\mu}$, i.e.,
%
\begin{eqnarray}
	\nabla_{\nu}
	(\hat{b}^{\mu} u^{\nu} - \hat{b}^{\nu} u^{\mu})
	& = &
	0
	, \label{eq:app:InductionEquationNonReconnecting} \\
	\nabla_{\nu}
	(\check{b}^{\mu} u^{\nu} - \check{b}^{\nu} u^{\mu})
	& = &
	- \frac{\check{b}^{\mu}}{c \tau_{\rm rec}}
	, \label{eq:app:InductionEquationDecayable}
\end{eqnarray}
where the parameter $\tau_{\rm rec}$ is the decay time-scale defined in the comoving frame.
These equations are the covariant formulation of the phenomenological magnetic dissipation.
Equations (30) and (31) of \citet{Drenkhahn02} are recovered for a (1) steady, (2) spherical symmetric, (3) pure radial flow with (4) pure toroidal magnetic fields for both $\hat{b}^{\mu}$ and $\check{b}^{\mu}$.
Note that all the right-hand side quantities in equations (30) and (31) of \citet{Drenkhahn02} are defined in the laboratory frame.

Our picture is different from the striped wind model (see section \ref{app:MagneticConversion}), and then we rewrite the `non-reconnecting toroidal' field $\hat{b}$ with $\bar{b}$ and the `decayable turbulent' one $\check{b}$ with $\delta b^{\mu}$ ($b^{\mu} = \bar{b}^{\mu} + \delta b^{\mu}$).
Adding equations (\ref{eq:app:InductionEquationNonReconnecting}) and (\ref{eq:app:InductionEquationDecayable}), we describe magnetic dissipation as
\begin{eqnarray}
	\nabla_{\nu}
	(b^{\mu} u^{\nu} - b^{\nu} u^{\mu})
	& = &
	- \frac{\delta b^{\mu}}{2 c \tau_{\rm diss}}
	, \label{eq:app:InductionEquationDissipation}
\end{eqnarray}
where we replace $\tau_{\rm rec}$ into $2 \tau_{\rm diss}$ for later convenience.
The induction equation for $\bar{b}^{\mu}$ is discussed in section \ref{app:MagneticConversion}.
Note that $\delta b^{\mu}$ is not toroidal unlike $\check{b}^{\mu}$ in \cite{Drenkhahn02}.

\subsection{Magnetic Conversion}\label{app:MagneticConversion}


\citet{Drenkhahn02} considered that the non-reconnecting fraction of the magnetic energy is constant of the system (equation (\ref{eq:app:InductionEquationNonReconnecting})).
However, in our picture, the non-reconnecting toroidal field is converted into the decaying turbulent field by MHD instabilities and then we extend the induction equation of $\bar{b}^{\mu}$ in the similar way of section \ref{app:MagneticDissipation} in order to allow the non-reconnecting fraction decreasing along the flow.
Introducing a sink term to the induction equation of $\bar{b}^{\mu}$, we obtain
\begin{eqnarray}
	\nabla_{\nu}
	(\bar{b}^{\mu} u^{\nu} - \bar{b}^{\nu} u^{\mu})
	& = &
	- \frac{\bar{b}^{\mu}}{2 c \tau_{\rm conv}}
	, \label{eq:app:InductionEquationConversion}
\end{eqnarray}
where $\tau_{\rm conv}$ is the comoving conversion time-scale and is an additional parameter of the system.
In term of the energy spectrum of the magnetic turbulence \citep[cf.][]{Zhou&Matthaeus90c}, $\tau_{\rm conv}$ determines the injection of the large-scale turbulence while $\tau_{\rm diss}$ determines dissipation of the small-scale turbulence \citep[see also section 4.3 of][]{Tanaka+18}.

\subsection{Derivation of Equations (\ref{eq:TotalEnergyConservation}) $-$ (\ref{eq:TurbulentMagneticEnergyConservation})}\label{app:Derivation}

Our non-ideal MHD system is fully described by replacing equations (\ref{eq:app:EnergyMomentumConservation}) and (\ref{eq:app:InductionEquation}) into equations (\ref{eq:app:EnergyMomentumConservationCooling}), (\ref{eq:app:InductionEquationDissipation}) and (\ref{eq:app:InductionEquationConversion}) with $b^{\mu} = \bar{b}^{\mu} + \delta b^{\mu}$ in the covariant form.
In order to obtain equations (\ref{eq:TotalEnergyConservation}) $-$ (\ref{eq:TurbulentMagneticEnergyConservation}), we impose the following assumptions: (i) steady ($\partial_t = 0$), (ii) cylindrical one-dimensional geometry ($g_{\mu \nu} = {\rm diag}(-1, 1, R^2, 1)$), (iii) velocity field along the jet axis ($u^{\mu} = (\gamma, 0, 0, u))$ and (iv) the non-reconnecting toroidal magnetic field ($\bar{b}^{\mu} = (0, 0, \bar{b}, 0))$.
(v) The decaying turbulent magnetic field, converted from the toroidal one, is set to be isotropic in the comoving frame ($\langle b_{\mu} b^{\mu} \rangle = \bar{b}^2 + \delta b^2$ and $\langle b_{\mu} \delta b^{\mu} \rangle = \delta b^2$) and satisfies $u_{\mu} \delta b^{\mu} = 0$ because $u_{\mu} b^{\mu} = u_{\mu} \bar{b}^{\mu} = 0$, where $\langle~\rangle$ represents the ensemble average.
The total magnetic field four-vector is $b^{\mu} = (u \delta b_z, \delta b_R, (\bar{b} + \delta b_{\phi})/ R, \gamma \delta b_z)$ with $\langle \delta b_i \delta b_j \rangle = (\delta b^2 / 3) \delta_{i j}$ ($i,j = R, \phi, z$).

Equation (\ref{eq:TotalEnergyConservation}) is the ensemble average of the time-component of equation (\ref{eq:app:EnergyMomentumConservationCooling}), i.e.,
\begin{eqnarray}
	\nabla_{\nu} \langle T^{t \nu} \rangle
 	& = &	\nabla_{\nu} \left( (w + \langle b^2 \rangle) u^t u^{\nu} + ( p + \langle b^2 \rangle ) / 2 ) g^{t \nu} - b^t b^{\nu} \right)
 	\nonumber \\
	& = &
	\frac{d}{d z} \left[\gamma u \left(w + \bar{b}^2 + \frac{2}{3} \delta b^2 \right) \right]
	= \gamma \frac{\Lambda_{\rm cool}}{c}
	. \label{eq:app:TotalEnergyConservation}
\end{eqnarray}
We treat $\delta b^{\mu}$ by contracting equations (\ref{eq:app:InductionEquationDissipation}) and (\ref{eq:app:InductionEquationConversion}) with $b_{\mu}$ and taking ensemble average.
We obtain
\begin{eqnarray}
	&  &\langle b_{\mu} \nabla_{\nu}
	(b^{\mu} u^{\nu} - b^{\nu} u^{\mu}) \rangle
    \nonumber \\
	& = &
    \langle b^2 \rangle \nabla_{\nu} u^{\nu} 
    + u^{\nu} \nabla_{\nu} \langle b^2 \rangle / 2 - \langle b_{\mu} b^{\nu} \rangle \nabla_{\nu} u^{\mu}	
    \nonumber \\
	& = &
    \langle b^2 \rangle \frac{d u}{d z} + u \frac{d}{d z} \frac{\langle b^2 \rangle}{2} - \langle b_{\mu} b^{\nu} \rangle \nabla_{\nu} u^{\mu}
    =
	- \frac{ \delta b^2 / 2}{c \tau_{\rm diss}}
	, \label{eq:app:TotalMagneticEnergyConservation}
\end{eqnarray}
and
\begin{eqnarray}
	& & \langle b_{\mu} \nabla_{\nu}
	(\bar{b}^{\mu} u^{\nu} - \bar{b}^{\nu} u^{\mu}) \rangle
    \nonumber \\
	& = &
    \bar{b}^2 \nabla_{\nu} u^{\nu} + u^{\nu} \nabla_{\nu} \bar{b}^2 / 2 - \bar{b}_{\mu} \bar{b}^{\nu} \nabla_{\nu} u^{\mu}	
    \nonumber \\
	& = &
    \bar{b}^2 \frac{d u}{d z} + u \frac{d}{d z} \frac{\bar{b}^2}{2}
    =
	- \frac{\bar{b}^2 / 2}{c \tau_{\rm conv}}
	, \label{eq:app:ToroidalMagneticEnergyConservation}
\end{eqnarray}
respectively.
Equation (\ref{eq:app:ToroidalMagneticEnergyConservation}) corresponds to equation (\ref{eq:ToroidalMagneticEnergyConservation}) and we obtain the equation for the turbulent magnetic field (equation (\ref{eq:TurbulentMagneticEnergyConservation})) from the difference between equations (\ref{eq:app:TotalMagneticEnergyConservation}) and (\ref{eq:app:ToroidalMagneticEnergyConservation}).
Finally, we contract equation (\ref{eq:app:EnergyMomentumConservationCooling}) with $u_{\mu}$ and take the ensemble average,
\begin{eqnarray}
	& & - u_{\mu} \nabla_{\nu} \langle T^{\mu \nu} \rangle
 	\nonumber \\
 	& = &
 	\nabla_{\nu} \left( u^{\nu} (w + \langle b^2 \rangle) \right) - u^{\nu} \nabla_{\nu} ( p + \langle b^2 \rangle  / 2 ) + u_{\mu} \langle b^{\nu} \nabla_{\nu} b^{\mu} \rangle
 	\nonumber \\
	& = &
	u \frac{d}{d z} \left( w - p + \frac{\langle b^2 \rangle}{2}  \right) + (w + \langle b^2 \rangle ) \frac{d u}{d z} + u_{\mu} \langle b^{\nu} \nabla_{\nu} b^{\mu} \rangle 
 	\nonumber \\
	& = &
	- \frac{\Lambda_{\rm cool}}{c}
	. \label{eq:app:TotalInternalEnergyConservation}
\end{eqnarray}
Combining equations (\ref{eq:app:TotalInternalEnergyConservation}) with (\ref{eq:app:TotalMagneticEnergyConservation}), we obtain equation (\ref{eq:InternalEnergyConservation}), where we use $u_{\mu} b^{\mu} = 0$.

\section{Asymptotic Solutions of Velocity Profile}\label{app:AsymptoticSolutions}

For the purpose of this section, the equation of state with the constant adiabatic index $\hat{\gamma}$ is adopted.
The relation between the pressure $p$ and the internal energy density $e_{\rm int}$ is set to
\begin{eqnarray}\label{eq:SimpleEOS_p}
	p
	& = &
	(\hat{\gamma} - 1) e_{\rm int}
\end{eqnarray}
and then the enthalpy density is
\begin{eqnarray}\label{eq:SimpleEOS_w}
	w
	& = &
	\rho c^2 + \frac{\hat{\gamma}}{\hat{\gamma} - 1} p,
\end{eqnarray}
where $\rho$ is the mass density.
In this appendix, we just need to replace $\hat{\Gamma}$ in the equations in section \ref{sec:Model} to $\hat{\gamma}$.

\subsection{Cooling}\label{app:CoolingAcceleration}

Here, we consider a pure hydrodynamic flow ($\bar{b} = \delta b = 0$) and then only the first term of the right-hand side of equation (\ref{eq:DifferentialEquationForVelocity}) remains.
For relativistically-hot ($p \gg \rho c^2$) ultra-relativistic flows ($u \approx \gamma \gg 1$ and $\beta \approx 1$), equation (\ref{eq:DifferentialEquationForVelocity}) becomes
\begin{eqnarray}\label{eq:CoolingAnalytic}
	u'
	& \approx &
	\frac{p}{w - \hat{\gamma} p}
	\nonumber \\
	& \approx &
	g^{-1} - g^{-2} \frac{\rho c^2}{p}
	~{\rm with}~
	g \equiv \hat{\gamma} \frac{2 - \hat{\gamma}}{\hat{\gamma} - 1},
\end{eqnarray}
where the normalized derivative operator $' \equiv x_{\rm cool} d / d x$ is introduced.
At the second line of equation (\ref{eq:CoolingAnalytic}), the right-hand side is expanded first-order in $\rho c^2 / p$.
Eliminating $\rho, p, e_{\rm int}$, and $w$ from equation (\ref{eq:InternalEnergyConservation}) with the use of equations (\ref{eq:MassFlux}), (\ref{eq:SimpleEOS_p}), (\ref{eq:SimpleEOS_w}), and (\ref{eq:CoolingAnalytic}), we obtain
\begin{eqnarray}
	(\hat{\gamma} - 1) u' (1 - g u') + g u u'' + 1 - g u' = 0.
\end{eqnarray}
The assumption of $u \gg 1$ gives $u u'' \approx (\hat{\gamma} - 1) u'^2$ and then we obtain $u \propto x^{1/\hat{\gamma}}$ from this differential equation.

Interestingly, cooling acceleration does work even in non-relativistic flows so that cooling acceleration is not a relativistic effect like the reduction of the `thermal' mass of the fluid by cooling.
In the non-relativistic limit for both the velocity ($u \approx \beta \ll 1$) and temperature ($w \beta^2 \approx \rho v^2$), equation (\ref{eq:DifferentialEquationForVelocity}) becomes
\begin{eqnarray}\label{eq:CoolingAnalyticNonRela}
	\beta'
	& \approx &
	\frac{p}{\rho v^2 - \hat{\gamma} p},
\end{eqnarray}
where $v = c \beta$.
The right-hand side of equation (\ref{eq:CoolingAnalyticNonRela}) is positive for supersonic flows $v > c_s$, where $c_s^2 \equiv \hat{\gamma} p / \rho$.
However, cooling acceleration beyond the sonic point can be significant only for relativistically-hot flows, because a supersonic flow is accelerated by extracting its thermal energy which is always smaller than the kinetic energy for non-relativistic supersonic flows.

\subsection{Conversion}\label{app:ConversionAcceleration}

Here, we consider the case that the second term of the right-hand side of equation (\ref{eq:DifferentialEquationForVelocity}) is important, i.e., $\Lambda_{\rm cool} = 0$ and $\tau_{\rm diss} \rightarrow \infty$.
For high-$\sigma$ ultra-relativistic flows ($u \approx \gamma \gg 1$ and $\beta \approx 1$), equation (\ref{eq:DifferentialEquationForVelocity}) becomes
\begin{eqnarray}\label{eq:ConversionAnalytic}
	u'
	& \approx &
	\frac{\bar{b}^2 / 3}{w - \hat{\gamma} p + (4/9) \delta b^2}
	\nonumber \\
	& \approx &
	\frac{3}{4 \chi},
	~{\rm with}~
	\chi \equiv \frac{\delta b^2}{\bar{b}^2},
\end{eqnarray}
where we set $\delta b^2 \gg w$ in the second line and the normalized derivative operator $' \equiv x_{\rm conv} d / d x$ is introduced.
Dividing equations (\ref{eq:ToroidalMagneticEnergyConservation}) and (\ref{eq:TurbulentMagneticEnergyConservation}) by $\bar{b}^2$ and $\delta b^2$, respectively and then subtracting the two equations from each other, we obtain
\begin{eqnarray}\label{eq:ConversionAnalytic2}
	\frac{\chi'}{\chi}
	=
	\frac{1}{u}
	+
	\frac{1}{u \chi}
	+
	\frac{2}{3} \frac{u'}{u}.
\end{eqnarray}
Eliminating $\chi$ from the equation (\ref{eq:ConversionAnalytic2}) with the use of equation (\ref{eq:ConversionAnalytic}), we find
\begin{eqnarray}
    u u'' + 2 u'^2 + u' = 0.
\end{eqnarray}
The assumption of $u \gg 1$ gives $u u'' \approx 2 u'^2$ and then we obtain $u \propto x^{1/3}$ from this differential equation and also obtain $\chi \propto x^{2/3}$ from equation (\ref{eq:ConversionAnalytic}).

\subsection{Dissipation}\label{app:DissipationAcceleration}

The last term of the right-hand side of equation (\ref{eq:DifferentialEquationForVelocity}) is less important than the other terms because a large value of $\delta b^2 / \tau_{\rm diss}$ immediately heating up plasma, i.e., $(4/3) - \hat{\Gamma} \rightarrow 0$.
Instead, we reexamine the case of direct dissipation of the toroidal magnetic field into the plasma heat \citep[cf.][]{Lyubarsky&Kirk01, Drenkhahn02, Zrake&Arons17}.
We recover the equations of direct dissipation of the toroidal magnetic field by setting the right-hand side of equation (\ref{eq:TurbulentMagneticEnergyConservation}) equals to zero, i.e., the turbulent magnetic field immediately dissipates rather than convected by the flow ($\bar{b}^2 / \tau_{\rm conv} = \delta b^2 / \tau_{\rm diss}$).
For high-$\sigma$ ultra-relativistic flows ($u \approx \gamma \gg 1$ and $\beta \approx 1$), equation (\ref{eq:DifferentialEquationForVelocity}) becomes
\begin{eqnarray}\label{eq:DissipationAnalytic}
	u'
	& \approx &
	\frac{2 - \hat{\gamma}}{2} \frac{\bar{b}^2}{w - \hat{\gamma} p}
	\nonumber \\
	& \approx &
	\left\{
		\begin{array}{ll}
			\displaystyle{\frac{1}{\hat{\gamma} \xi}} & (p \gg \rho c^2) \\
			\displaystyle{\frac{2 - \hat{\gamma}}{2 \eta}} & (w \approx \rho c^2)
 		\end{array}
	\right.
\end{eqnarray}
where we introduced $\xi \equiv p / \bar{b}^2, \eta \equiv \rho c^2 / \bar{b}^2$ and the normalized derivative operator $' \equiv x_{\rm diss} d / d x$ is introduced.

For cylindrical jets, plasma would be heated up to a relativistic temperature by magnetic dissipation, i.e., $p \gg \rho c^2$.
Combining equations (\ref{eq:InternalEnergyConservation}) and (\ref{eq:ToroidalMagneticEnergyConservation}), we obtain, with the use of equations (\ref{eq:MassFlux}) and (\ref{eq:SimpleEOS_w}),
\begin{eqnarray}\label{eq:DissipationAnalytic2}
	\frac{\xi'}{\xi}
	=
	\frac{1}{u}
	+
	\frac{1}{u \xi}
	+
	(2 - \hat{\gamma}) \frac{u'}{u}.
\end{eqnarray}
Eliminating $\xi$ from equation (\ref{eq:DissipationAnalytic2}) with the use of equation (\ref{eq:DissipationAnalytic}), we find
\begin{eqnarray}
    u u'' + 2 u'^2 + u' = 0.
\end{eqnarray}
The assumption of $u \gg 1$ gives $u u'' \approx 2 u'^2$ and then we obtain $u \propto x^{1/3}$ which is the same as the case of conversion acceleration (appendix \ref{app:ConversionAcceleration}).

For spherical geometry, on the other hand, plasma is not heated up to a relativistic temperature as a result of flow expansion.
In order to recover the result of \citet{Drenkhahn02}, we consider the cold limit $w \approx \rho c^2$ (omitting equation (\ref{eq:InternalEnergyConservation}).
Combining equations (\ref{eq:MassFlux}) and (\ref{eq:ToroidalMagneticEnergyConservation}), we obtain
\begin{eqnarray}\label{eq:DissipationAnalytic3}
	\frac{\eta'}{\eta}
	=
	\frac{1}{u}
	+
	\frac{u'}{u}.
\end{eqnarray}
Eliminating $\eta$ from equation (\ref{eq:DissipationAnalytic3}) with the use of equation (\ref{eq:DissipationAnalytic}), we find
\begin{eqnarray}
	u u'' + u'^2 + u' = 0.
\end{eqnarray}
The assumption of $u \gg 1$ gives $u u'' \approx u'^2$ and then we obtain $u \propto x^{1/2}$ which is different from the case in the relativistically-hot limit.
$u \propto x^{1/2}$ is also different from the result of \citet{Drenkhahn02} ($u \propto r^{1/3}$) because they set $\tau_{\rm diss}$ is not constant but $\tau_{\rm diss} \propto \gamma$.
Taking into account $\tau_{\rm diss} \propto \gamma \approx u$, we obtain the same result $u \propto x^{1/3}$ as \citet{Drenkhahn02} with the same procedure as above.
It should be noted that $u \propto x^{1/3}$ is obtained without the expansion acceleration term, i.e., the plasma heating by magnetic dissipation and subsequent expansion acceleration are not important for dissipation acceleration.
Spherical geometry just ensures the flow to be cold by adiabatic expansion.

\section{Profiles of various inlet boundary}\label{app:VariousInletBoundary}

%
\begin{figure*}
	\includegraphics[width=2.0\columnwidth]{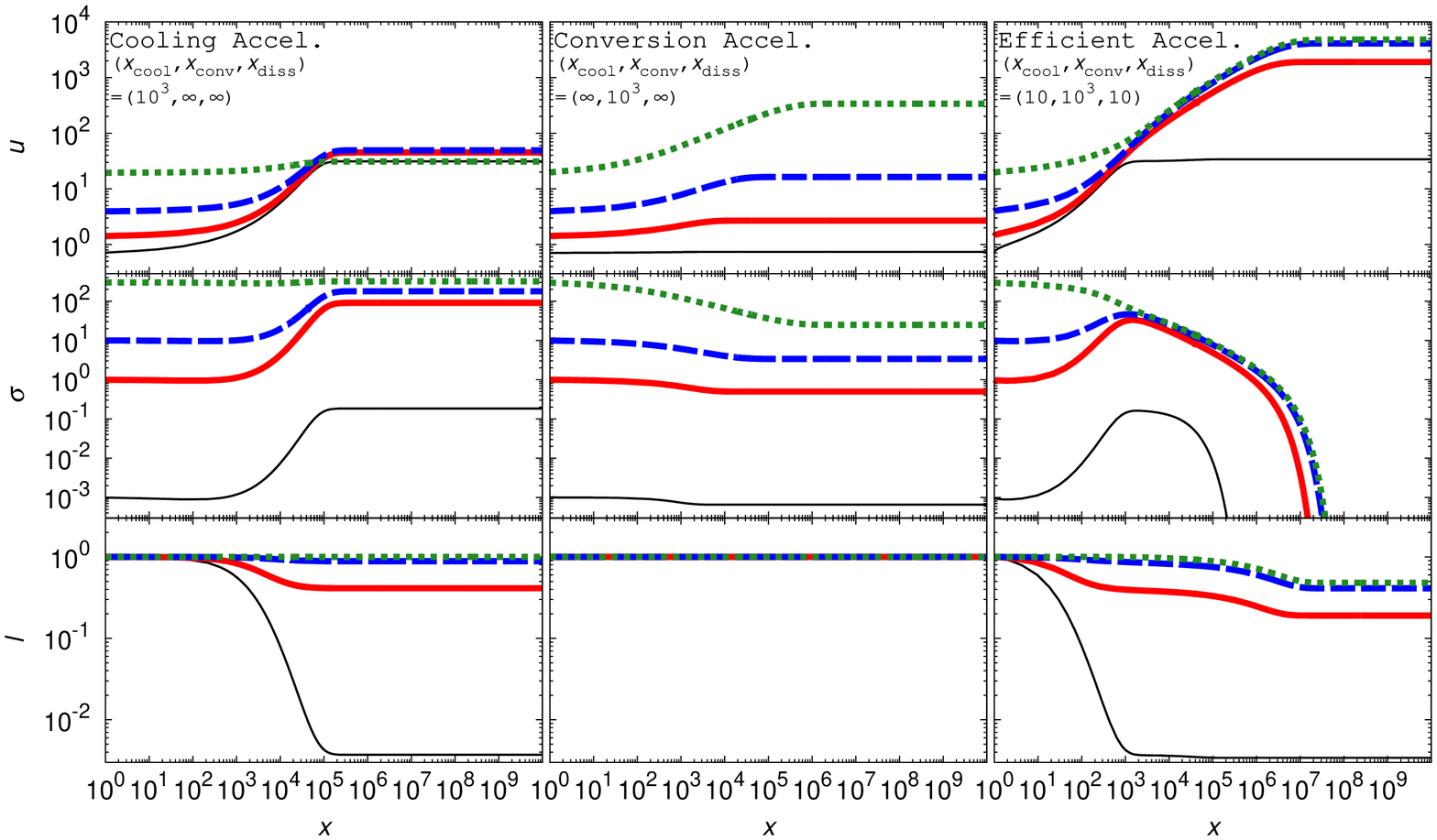}
    \caption{
    The profiles of the four velocity $u(x)$ (top row), magnetization $\sigma(x)$ (middle row) and the total energy flux $l(x)$ (bottom row) for the different inlet boundary conditions $\sigma_0 = 10^{-3}$ (black thin line), $1$ (red solid line), $10$ (blue dashed line), and $3 \times 10^{2}$ (green dotted line).
    The left three panels consider the case of cooling acceleration $(x_{\rm cool},x_{\rm conv},x_{\rm diss})=(10^3,\infty,\infty)$ (section \ref{sec:CoolingAcceleration}).
    The middle three panels consider the case of conversion acceleration $(x_{\rm cool},x_{\rm conv},x_{\rm diss})=(\infty,10^3,\infty)$ (section \ref{sec:ConversionAcceleration}).
    The right three panels consider the case of efficient acceleration $(x_{\rm cool},x_{\rm conv},x_{\rm diss})=(10,10^3,10)$ (section \ref{sec:EfficientAcceleration}).
    }
    \label{fig:SgmDependence}
\end{figure*}

Fig. \ref{fig:SgmDependence} shows the flow profiles for a various $\sigma_0$.
The characteristic lengths are fixed to $(x_{\rm cool},x_{\rm conv},x_{\rm diss})=(10^3,\infty,\infty)$ for cooling acceleration (left column), $(x_{\rm cool},x_{\rm conv},x_{\rm diss})=(\infty,10^3,\infty)$ for conversion acceleration (middle column) and $(x_{\rm cool},x_{\rm conv},x_{\rm diss})=(10,10^3,10)$ for efficient acceleration (right column).
The inlet boundary conditions are $\sigma_0 = 10^{-3}$ (black thin), $1$ (red solid), $10$ (blue dashed), and $3 \times 10^{2}$ (green dotted), where $\sigma_0 \lesssim \gamma_{\rm max}^{2/3} \approx 460$ for $\gamma_{\rm max} = 10^4$ from the requirement $\Theta_0 > 0$ (see section \ref{sec:NumericalImplementation}).

For cooling acceleration, $u_{\infty}$ are similar to each other, while $\sigma_{\infty}$ and $l_{\infty}$ are different for different $\sigma_0$.
Cooling acceleration is ineffective for the high-$\sigma_0$ cases and then $u_0 \approx u_{\infty}$.
The inlet internal energy density is $e_{\rm int,0} \approx$ 0.86 and 0.84 for $\sigma_0 = 10^{-3}$ and $10^{-2}$, respectively so that the velocity profiles for $\sigma_0 \ll 1$ are almost the same.

Conversion acceleration is significant for the high-$\sigma_0$ flows and inefficient for $\sigma_0 \le 1$.
On the other hand, for efficient acceleration, the terminal velocity is $u_{\infty} \approx \gamma_{\rm max} / 5$ even for $\sigma_0 = 1$.
For efficient acceleration, both the velocity and magnetization profiles are similar to each other for $\sigma_0 \ge 1$ beyond $x > 10^3$.


\bsp	
\label{lastpage}
\end{document}